\def\diag{{\rm diag}}\def\dim{{\rm dim}}
\def\det{{\rm det}}\def\tr{{\rm Tr}}
\def \ak{$A_{k-1}$}\def \dk{$D_{k+2}$}
\def\cC{{\cal C}}\def\cF{{\cal F}}\def\cG{{\cal G}}\def\cH{{\cal H}}
\def\cL{{\cal L}}\def\cM{{\cal M}}\def\cN{{\cal N}}\def\cO{{\cal O}}
\def\cP{{\cal P}}\def\cQ{{\cal Q}}\def\cR{{\cal R}}
\def\cS{{\cal S}}\def\cZ{{\cal Z}}
\def\quat{{\bf H}}
\def\eqdef{\stackrel{\rm def}{=}}
\def\twomat#1#2#3#4{\left(\matrix{#1&#2\cr #3&#4\cr}\right)}

\def\twovec#1#2{\left(\matrix{#1\cr #2\cr}\right)}
\def\pinst#1#2{#1\choose #2}
\def\sp#1#2{{#1\brack#2}}

\def\o#1#2{{#1\over#2}}
\def\bz{\bar z}
\def\um{{1 \over 2}}
\def\un{{\bf 1}}
\def\bx{{\bf X}}
\def\G{\cG}
\def\r{{\bf R}}
\def\c{{\bf C}}

\def\csg{\c^2/\Gamma}

\documentstyle[12pt]{article}
\begin{document}
\newcommand{\hika}{HyperK\"ahler~}
\newcommand{\ka}{K\"ahler~}
\newcommand{\CY}{Calabi-Yau~}
\newcommand {\cJ}[1]{{\cal J}^{#1}}
\newcommand{\cJt}[3]{{\cal J}^{#1}_{#2#3}}
\newcommand{\epsi}[3]{\varepsilon^{#1#2#3}}
\begin{titlepage}
\begin{flushright}
SISSA/44/92/EP\\
IFUM/443/FT
\end{flushright}
\vskip 0.8cm
\begin{center}
{\Large\bf
ALE Manifolds and Conformal Field Theories}
\end{center}
\vskip 0.2cm
\begin{center}
Damiano Anselmi, Marco Bill\'o, Pietro Fr\'e
\end{center}
\begin{center}
{\it SISSA - International School for Advanced Studies, via Beirut 2-4,
34140 Trieste, Italy \\ and I.N.F.N. - Sezione di Trieste - Trieste, Italy}
\end{center}
\vskip 0.2cm
\begin{center}
Luciano Girardello
\end{center}
\begin{center}
{\it Dipartimento di Fisica, Universit\`a di Milano, via Celoria 6,
20133 Milano, Italy \\ and I.N.F.N. - Sezione di Milano, Milano, Italy}
\end{center}
\vskip 0.2cm
\begin{center}
Alberto Zaffaroni
\end{center}
\begin{center}
{\it SISSA - International School for Advanced Studies, via Beirut 2-4,
34140 Trieste, Italy \\ and I.N.F.N. - Sezione di Trieste - Trieste,
Italy}
\end{center}
\vskip 0.3cm
\begin{center}
{\bf Abstract}
\end{center}
\vskip 0.1cm\noindent
We address the problem of constructing the family of (4,4)
theories associated with the sigma-model on a parametrized family
${\cal M}_{\zeta}$ of Asymptotically Locally Euclidean (ALE)
manifolds. We rely on the ADE classification of these manifolds and on
their construction as HyperK\"ahler quotients, due to Kronheimer.
 So doing  we are able to define the family of
(4,4) theories corresponding to a ${\cal M}_{\zeta}$
family of ALE manifolds as the deformation of a solvable orbifold
${\bf C}^2 \, / \, \Gamma$ conformal field-theory, $\Gamma$ being a
Kleinian group.
We discuss the relation among the algebraic structure underlying the
topological and metric properties of self-dual 4-manifolds
and the algebraic properties of non-rational
(4,4)-theories admitting an infinite spectrum of primary fields.
In particular, we identify the Hirzebruch
signature $\tau$ with the dimension of the local polynomial ring
${\cal R}=\o {{\bf C}[x,y,z]}{\partial W}$ associated with the ADE
singularity, with the number of non-trivial conjugacy classes in the
corresponding Kleinian group and with the number of short
representations of the (4,4)-theory minus four.
\vskip 0.2cm
\hrule
\begin{flushleft}
hep-th/9304135
\end{flushleft}
\end{titlepage}
\section{Introduction}
\label{intro}
Recently a considerable amount of efforts have been devoted to the
construction of exact conformal
 field theories describing superstring propagation on non-trivial
 space-time geometries
\cite{xspacecftgeneral,blackholes,callan,previous,newkounnas}.
One research-line has focused
on the construction of Black-hole-like solutions while  another
one aimed at the search
for gravitational-instanton-like backgrounds.
Following the pioneering work of \cite{callan}
some of us have recently  established  the following result \cite{previous} :
\par
{\it Stringy gravitational instantons can be identified with $(4,4)$ $c=6$
conformal
field-theories in the very same way as $(2,2)$ $c=9$ theories can be
identified with Calabi-Yau three-folds }
\par
In particular  in \cite{previous} we have  introduced a generalization of
the notion of
HypeK\"ahler manifolds, that incorporates also the torsion  background
and that defines the
most general conditions under which a $(1,1)$-supersymmetric
$\sigma$-model develops a (4,4)-global
supersymmetry, leading to a (4,4) superconformal field theory.
\par
Central to the analysis of \cite{previous} is the concept of
abstract Hodge-diamond.
Let us briefly  review it.  We start by treating the $(4,4)$ $c=6$ theory as
a $(2,2)$ one (something that we can always do) and we denote the (2,2)
primary fields by
$\Psi {\pinst{h,{\tilde h}}{q,{\tilde q}}}(z,\bz)$, where $h,
{\tilde h}$are the left (right) conformal weights and $q,{\tilde q}$ are
the left (right) $U(1)$-charges. In the $(4,4)$ language the primary fields
are instead denoted by $\Psi\sp{h,{\tilde h}}{J,{\tilde
J}}^{m,\tilde m}(z,\bz)$ where
$h,{\tilde h}$ are the left (right) conformal weights, $J,{\tilde J}$ the
left (right) isospins and $m,{\tilde m}$ their third components.
The $(4,4)$ representations put together  the $(2,2)$
primary fields in bigger multiplets. In particular the relation between
the $U(1)$ charges ${q,{\tilde q}}$ and the isospin third components is
$q=2m$ (${\tilde q}=2{\tilde m}$). This happens because:
\begin{eqnarray}
{ \varphi}(z)&=&\o{1}{\sqrt{2}} \tau (z)\nonumber\\
{\tilde \varphi}(\bz)&=&\o{1}{\sqrt{2}}{\tilde \tau} (\bz)
\label{introd2}
\end{eqnarray}
where ${ \varphi}(z)$ is the free bosonic field that bosonizes the
left $U(1)$ current $j(z)=\partial\varphi(z)$ while $\tau(z)$ is the field
that bosonizes the level one $SU(2)$ algebra
contained in the $c=6,N=4$ algebra:
$A_3 (z) = \partial \tau(z)$,
$ A^{\pm}(z)={\rm exp}\left (\pm \tau(z) \right )$
(similarly for the right sector).
Using these notations we can attach to a $(2,2)$ $c=6$ theory
with integral $U(1)$ charges
an array of numbers (the abstract Hodge-diamond)
\begin{eqnarray}
\label{introd1}
\matrix{~&~&h^{(0,0)}&~&~&\cr
{}~&h^{(1,0)}&~&h^{(0,1)}&~&\cr
h^{(2,0)}&~&h^{(1,1)}&~&h^{(0,2)}&\cr
{}~&h^{(1,2)}&~&h^{(2,1)}&~&\cr
{}~&~&h^{(2,2)}&~&~&\cr}
\end{eqnarray}
where $h^{(p,q)}$ denotes the number of chiral-chiral primary fields
$\Psi {\pinst{p/2\, , \,q/2}{p \, , \, q}}(z,\bz)$ ($a=1,....,h^{(p,q)}$).
These fields span linear spaces that we name ${\cal H}^{(p,q)}$, the
abstract Dolbeault cohomology groups. We recall that primary chiral
fields are those admitting a regular operator product expansion with
one of the supercurrents ($G^{+}$ in this case) and are characterized by
the equation $q=2h$ (primary antichiral fields have regular operator
product with the other supercurrent ($G^{-}$) and are characterized
by $q=-2h$). In a unitary (not necessarily non-degenerate \cite{lercheetal})
(2,2)-theory there is a bound on the maximal $U(1)$-charge of primary
chirals: $q_{max}\, = \, \o{c}{3} \, \Rightarrow \, h_{max} \, = \,
\o{c}{6}$. Actually both in the left and right sectors there is a unique
top chiral primary $\rho {\pinst{\o{c}{6},0}{\o{c}{3} ,0}}(z,\bz)$ and
${\tilde \rho}{\pinst{0,\o{c}{6}}{0,\o{c}{3} }}(z,\bz)$.
Furthermore another important general property of $(2,2)$-theories,
namely the spectral flow,  allows to relate different chiral primaries.
In the $c=6$ case, by means of the operations
\begin{eqnarray}
U_{(1,0)} \, : \, e^{ i \sqrt{\o{c}{3} }\varphi (z)} \,
\Psi{\pinst{\o p2\, , \,\o q2}{p \, , \, q}}(z,\bz) \,
?& = & \,
\Psi{\pinst{\o p2 +1 \, , \,\o q2}{p +2\, , \, q}}(z,\bz)\nonumber\\
U_{(0,1)} \, : \, e^{i \sqrt{\o{c}{3} }\tilde\varphi (\bz )}
\Psi{\pinst{\o p2\, , \,\o q2}{p \, , \, q}}(z,\bz) \, & = & \,
\,\Psi{\pinst{\o p2 , \,\o q2+1}{p \, , \, q+2}}(z,\bz) \nonumber\\
U_{(1,1)} \, : \, e^{i \sqrt{\o{c}{3} } \, \left (
\varphi \left (z \right ) +{\tilde \varphi} \left (\bz \right )\right )}
\Psi {\pinst{\o p2\, , \,\o q2}{p \, , \, q}}(z,\bz) \,& = & \,
\Psi{\pinst{\o p2 +1\, , \,\o q2 +1}{p +2\, , \, q+2}}(z,\bz)
\label{introdspectralflow}
\end{eqnarray}
we realize the following isomorphisms ${\cal H}^{(p,q)} \, \approx \,
{\cal H}^{(p+2,q)} \, \approx \, {\cal H}^{(p+2,q+2)}$ that, together with
the previous information on the existence of a unique top chiral
primary imply the following structure for the abstract Hodge diamond
of the most general $(2,2)$  $c=6$ theory:
\begin{eqnarray}
\label{introdhodgediamond}
\matrix{~&~&1&~&~&\cr
{}~&h^{(1,0)}&~&h^{(1,0)}&~&\cr
1&~&h^{(1,1)}&~&1&\cr
{}~&h^{(1,0)}&~&h^{(1,0)}&~&\cr
{}~&~&1&~&~&\cr}
\end{eqnarray}
regardless whether the
theory is rational (with a finite number of primaries) or not (with an infinite
number of primaries as those associated with non compact target
manifolds).
In the (4,4) language the crucial $h^{(1,1)}$ number
just counts the short representations, whose lowest components are
the fields
$\Psi_{A} \sp{\um,\um}{\um,\um}^{m,{\tilde m}}(z,\bz)$, while
$h^{(1,0)}$ counts the short representations
$\Psi_{A} \sp{\um,0}{\um,0}^{m}(z,\bz)$, if present.
In the case where the target space of the $\sigma$-model  is a compact
torsionless manifold the abstract Hodge diamond is nothing else but
the geometrical Hodge diamond,
displaying the dimensions of the Dolbeault cohomology
groups $H^{(p,q)}$ of harmonic (p,q)-forms.
So in the case of the  rational (4,4)-theories associated
with the $K_3$-manifold and with the torus $T^4$ we have:
\begin{equation}
\label{introd3-4}
\begin{array}{l}{\underline K_3}\\ \null \\ \null \\ \null \\ \null \end{array}
\hskip 0.3cm
\matrix{~&~&1&~&~&\cr
{}~&0&~&0&~&\cr
1&~&20&~&1&\cr
{}~&0&~&0&~&\cr
{}~&~&1&~&~&\cr}
\hskip 0.5cm ; \hskip 0.5cm
\begin{array}{l}{\underline T_4}\\ \null \\ \null \\ \null \\ \null\end{array}
\hskip 0.3cm
\matrix{~&~&1&~&~&\cr
{}~&2&~&2&~&\cr
1&~&4&~&1&\cr
{}~&2&~&2&~&\cr
{}~&~&1&~&~&\cr}
\end{equation}
In the case where the (4,4) theory under consideration corresponds
 to a $\sigma$-model on a non-compact HyperK\"ahler,
or generalized HyperK\"ahler manifold, the geometrical interpretation
of the abstract Hodge diamond
is more involved.  Yet, as we stressed in \cite{previous},
it still provides the correct counting
for the zero-energy excitations of the light particles moving in the
instanton background. Indeed
we showed how to construct the emission vertices of all
such excitations in terms of representations of the abstract (4,4) theory.
In particular
we obtained
the following formulae for  the zero-mode counting:
\begin{eqnarray}
{\# } ~{\rm graviton~zero~modes}&=&3\, \left ( \, h^{(1,1)} \,
- \,1\, \right ) \, + \, 1 \nonumber\\
{\# } ~{\rm axion~zero~modes}&=& h^{(1,1)} \, + \, 2 \nonumber\\
{\# } ~{\rm gravitino~zero~modes}&=&
2 \, h^{(1,1)} \, + \, 4 \, h^{(1,0)}.
\label{introd2bis}
\end{eqnarray}
An explicit example studied in \cite{previous} is given by the
$SU(2) \, \otimes \, {\bf R}$ instanton of \cite{callan},
\cite{kounnasporrati}, \cite{dauriaregge},
and \cite{ivanov}, where the (4,4)-theory is constructed in terms of an $SU(2)$
current algebra at level $k$ and a supersymmetric Feigin-Fuchs boson.
This model has an untwisted sector where the
abstract Hodge-diamond is the same as the abstract Hodge diamond of
the $T^4$-torus recalled in eq. \ref{introd3-4}.
Shortly after \cite{previous} Kounnas et al \cite{newkounnas} have
shown that the same model has  also a twisted sector containing
additional short-representations, so that the abstract
Hodge-diamond has a richer structure that will be shortly discussed
in section \ref{su2xr}.

In the present paper we want to deepen our understanding of
the algebraic structure of (4,4)
theories in the non compact case, which is the one relevant
to gravitational instanton physics.
{}From the geometrical point of view, the essential novelty,
with respect to the compact case,
is the presence of a boundary at infinity that
forces upon us the distinction between absolute $H^{(p,q)}({\cal M})$
and relative $H^{(p,q)}({\cal M}, \partial{\cal M} )$ cohomology groups.
We have therefore to distinguish among the harmonic
$(p,q)$-forms those that have compact
support and those that are not normalizable. Only the first
type of forms correspond to true
deformations of the underlying HyperK\"ahler (or generalized HyperK\"ahler)
structure, although
all of them are associated with emission vertices of zero-energy states.
Indeed we just retrieve in a different set up the usual
distinction between the discrete and continuous part of the spectrum.
The abstract Hodge-diamond of the (4,4) theory encodes
the dimensions of the absolute cohomology groups;
however,  one can also write the corresponding relative
Hodge-diamond,  listing the number of short
representations that correspond to normalizable states
and hence to true moduli of the superconformal theory.
\par
To illustrate these ideas we investigate the problem
of constructing the (4,4)-theories
corresponding to the {\it Asymptotically Locally Euclidean
(ALE) self-dual four-manifolds}.
These gravitational instantons were extensively studied
in the late seventies by physicists and mathematicians
\cite{inst,gibbonshawking,eguchihanson,hawkingpope,hitchin}
and their classification,
already conjectured at that time, has been finally proved in
recent years by  Kronheimer
\cite{kro1,kro2}, who has also provided an algorithm for
their direct construction in terms of HyperK\"ahler quotients.
The algorithm in question,  reviewed in the present
paper in a language accessible to physicists,
relies heavily on the point of view that regards an ALE manifold as
the minimal resolution of singularities of an orbifold
${\bf C}^2/\Gamma$, where
${\bf C}^2 \, \approx \, \r^4$ is the usual Euclidean flat
space and $\Gamma$ is a finite Kleinian
subgroup of $SU(2)$. This viewpoint, that traces back the
properties of an ALE manifold to the
structure of its boundary at infinity (a three sphere
modded out by the Kleinian group $\Gamma$)
is of great value in relation with conformal field theories.
Indeed, precisely as it happens for the case of the
compactified internal dimensions, a superstring
propagates nicely on a singular
orbifold variety, the corresponding conformal field
theory being solvable and completely
determined \cite{dixonharveywitten}. Hence, just as orbifold theories
(see \cite{lecdixon} for a review)
have been a nice substitute for the (2,2)
conformal field theories of smooth Calabi-Yau manifolds, in the same way,
we retrieve all the essential properties of the  (4,4)
theories of ALE manifolds, by studying
the ${\bf C}^2/\Gamma$ orbifold theory. The key difference
with the compact case is that now we
have a point-group but no space-group, so that there
are twisted sectors but no lattice quantization
of the momenta.  The twist fields are the essential
ingredients in the construction of the short
representations, so that, at the level of the (4,4) theory,
the identification of the number of moduli
deformations with the number of conjugacy classes of the
corresponding Kleinian group becomes particularly natural.
\par
Geometrically the ALE manifolds can be described  as  affine
complex varieties in ${\bf C}^3$,
namely the zero loci of certain polynomial constraints
\begin{equation}
\{ \, x, \, y, \, z \} \, \in \, {\cal M}_{\Gamma} \, \left
( \, t_1, ..., t_{r} \, \right)
{}~~\longrightarrow~{\tilde W}_{\Gamma} \, \left ( \, x, \, y, \, z, \,
; \, t_1, \, ....\, , t_r \, \right )~=~0
\label{introd5},
\end{equation}
that are determined by the algebraic structure of the
Kleinian group $\Gamma$ and that depend
on $r$ complex parameters $\{ t_i \} \, (i=1,...,r)$,
$r$ being the number of non-trivial conjugacy
classes in $\Gamma$. In the limit $t_i \, = \, 0 $ the
locus (\ref{introd5}) reduces to the
orbifold ${\bf C}^2/\Gamma$ and for these values of
the moduli the (4,4)-theory is known and
solvable. We have not found, so far, any algorithm to construct
the (4,4)  theory corresponding to a generic point in moduli-space
and probably such an algorithm does not exist. Our station
with ALE manifolds is therefore the same as with Calabi-Yau
manifolds: we know the general properties of the associated (4,4)
CFT (respectively (2,2) theory) and we can solve it in some
special point of moduli space (the orbifold limit for ALE manifolds,
the orbifold limit or the
Gepner tensor product point for Calabi-Yau n-folds).
Generic points can be reached perturbatively by deforming around
the solvable point. In the ALE case this perturbation has a
particular nice geometrical interpretation: all the moduli are
twisted and they are the parameters associated with the
resolution of singularities.
\par
The whole point of our paper is to show how the algebraic
structure underlying the
topological and metric properties of self-dual 4-manifolds
(an instance of non-compact  Calabi-Yau spaces) is naturally matched
with the algebraic properties of non-rational
(4,4)-theories admitting an infinite spectrum of primary fields.
In \cite{previous} we have formulated the general scheme
for the abstract description of a gravitational instanton
as a left-right symmetric N=4 theory
and we have considered the example of the generalized HyperK\"ahler ,
non asymptotically flat $SU(2) \, \otimes \, {\bf R}$ instanton.
As remarked there, this instanton is
a point in a moduli space whose other points correspond to so
far unknown manifolds.
In the present publication we compare the general superconformal
framework with the case of the conventional HyperK\"ahler ALE
instantons whose moduli space is fully
understood in geometrical terms.
\par
Our paper is organized as follows. In section \ref{generale} we review
the definition of ALE manifolds, discuss their cohomology and point out
their classification in terms of the Kleinian groups acting on their
boundary. The essential algebraic information about Kleinian groups are
collected in section \ref{grouptheory}.
Next, relying on the structure of the above groups, we summarize
the counting of parameters for ALE metrics.
Section \ref{construction} recalls the main ideas of HyperK\"ahler quotients
and explains in a language accessible to physicists the main steps
of Kronheimer construction, emphasizing its relevance
for the errand of the conformal field theory derivation.
In particular we obtain
the precise mapping between the parameters of the HyperK\"ahler quotient
utilized by Kronheimer and the deformation parameters $t_i$ of the potential
${\tilde W}(x,y,z~;~t_i)$ appearing in eq.(\ref{introd5}).
Then we turn to the construction
of the ${\bf C}^2/\Gamma$ orbifold conformal field-theory. This
is described in section \ref{cft} where the structure of
the N=4 representations is also discussed, retrieving
both the short representations associated with compact support
cohomology and those associated with non-normalizable states.
In section \ref{part} we study the structure
of the partition function and we analyse it in characters of the N=4
theory. Finally in section \ref{su2xr} we make some concluding
remarks comparing the case we have considered with the complete structure
of the $SU(2) \, \otimes \,
{\bf R}$ instanton as it appears after the new results obtained by
Kounnas et al.
\cite{newkounnas}, and pointing out the open problems in the
interpretation of the chiral ring $\c[x,y,z]/\partial W$ multiplicative
structure in the context of CFT.

\section{Generalities on ALE manifolds}\label{generale}
\underline{\sl Non-compact \hika four-manifolds}
\vskip 0.2cm
On a four-dimensional \hika manifold $\cM$ there exist three covariantly
constant complex structures $\cJ i: T\cM\rightarrow T\cM$, $i=1,2,3$.
The metric is hermitean with respect to all of them and they satisfy the
quaternionic algebra: $\cJ i \cJ j = - \delta^{ij} + \epsi ijk \cJ k$.
In a vierbein basis $\{V^a\}$, the matrices $\cJt iab$ are
antisymmetric (by hermiticity). By covariant constancy, the three \hika
two-forms $\Omega^i=\cJt iab V^a\wedge V^b$ are closed: $d\Omega^i=0$.
Because of the quaternionic algebra constraint, the $\cJt iab$ can only
be either selfdual or
antiselfdual; we take them to be antiselfdual: $\cJt iab =-{1\over 2}
\epsilon_{abcd} \cJt icd$. Then the integrability condition for the
covariant constancy of $\cJ i$ forces the curvature two-form $R^{ab}$ to
be selfdual (thus automatically solving the vacuum Einstein equations).

A \hika manifold is in particular a \ka manifold with respect to each of
its complex structures. Choose one of the structures (say $\cJ 3$)
and fix a frame on $\cM$ well-adapted to it. Consider then the
Dolbeaut cohomology groups $H^{p,q}(\cM)$, of dimensions $h^{p,q}$. Since
$\cM$ is Ricci-flat, its first Chern class vanishes: $c_1(\cM)=0$; $\cM$ is a
(non-compact) \CY manifold and therefore $h^{2,0}=h^{0,2}=1$. It is easy
to see that $\Omega^{\pm}=\Omega^1\pm i \Omega^2$ are holomorphic (resp.
antiholomorphic) so that $[\Omega^+]\equiv H^{2,0}(\cM)$,
$[\Omega^-]\equiv H^{0,2}(\cM)$, where by $[\Omega^{\pm}]$ we mean the
cohomology classes of $\Omega^{\pm}$. $\Omega^3$ is the \ka form on $\cM$
and $[\Omega^3]$ is just one of the elements of $H^{1,1}(\cM)$.

On a non-compact manifold it is worth considering the ``compact-support''
cohomology groups, that coincide with the relative cohomology groups of
forms vanishing on the boundary at infinity of the manifold:
\[
H^p_{\bf c}= {\{{\bf L}^2\,{\rm integrable,\hspace{2pt} closed}\hspace{2pt}
 p-{\rm forms}\} \over \{{\bf L}^2\,{\rm integrable,\hspace{2pt} exact}
\hspace{2pt} p-{\rm forms}\}}=H^p(\cM,\partial\cM),
\]
of dimensions $b^p_{\bf c}$. Analogously we will consider the
compact support Dolbeaut cohomology
groups $H^{p,q}_{\bf c}$, of dimensions $h^{p,q}_{\bf c}$.
The Poincar\'e duality provides an isomorphism  $H_p(\cM)\sim H^{4-p}_{\bf
c}(\cM)$, where $H_p(\cM)$ are the homology groups. Call $b_p$ their
dimensions (the {\it Betti numbers}); then $b_p=b^{4-p}_{\bf c}$.

The fundamental topological invariants characterizing the gravitational
instantons were recognized long time ago (\cite{inst}, for a review see
\cite{ehphysrep}) to be the Euler
characteristic $\chi$ and the Hirzebruch signature $\tau$ of the base
manifold.

The Euler characteristic is the alternating sum of the Betti numbers:
\begin{equation}
\chi=\sum_{p=0}^{4}(-1)^p b_p=\sum_{p=0}^{4}(-1)^p b^{4-p}_{\bf c}=
\sum_{p=0}^{4}(-1)^p b^p_{\bf c}.\label{Euchar}
\end{equation}
The Hirzebruch signature is the difference between the number of
positive and negative eigenvalues of the quadratic form on $H^2_{\bf
c}(\cM)$ given by the cup product $\int_{\cM} \alpha\wedge\beta$, with
$\alpha,\beta\in H^2_{\bf c}(\cM)$. That is, if $b^{2(+)}_{\bf c}$ and
$b^{2(-)}_{\bf c}$ are the number of selfdual and anti-selfdual 2-forms
with compact support, $\tau=b^{2(+)}_{\bf c}-b^{2(-)}_{\bf c}$.
At this point, we need two observations.
\begin{enumerate}
\item
The \hika forms $\Omega^3, \Omega^{\pm}$, being covariantly
constant, cannot be ${\bf L}^2$ if the space is non-compact
\item
In the compact case they are the unique antiselfdual
2-forms, so that $b^{2(-)}=3$, $b^{2(+)}=\tau +3$.
Indeed from the expression of the Hirzebruch signature in terms of the
Hodge numbers,
\hspace{3pt} $\tau=\sum_{p+q=0{\rm mod}2}(-1)^p h^{p,q}$ (see
\cite[chap. 0, sec. 7]{grif}), using the consequences
of the Calabi-Yau condition
$c_1(\cM)=0\Rightarrow h^{2,0}=h^{0,2}=1$ and the fact that
$h^{0,0}=h^{2,2}=1$ we obtain $h^{1,1}=\tau +4$.
Hence the cohomology in degree two splits
as follows:
\[
\begin{array}{ccc}
h^{2,0} & h^{1,1} & h^{0,2} \\
1 & 1+(\tau +3) & 1
\end{array}\]
This leads to the conclusion that $\Omega^3\in H^{1,1}$ and
$\Omega^{\pm}\in H^{2,0}$ (resp. $H^{0,2}$) are the unique antiselfdual
two-forms.
\end{enumerate}
In the non compact case, by the observation (1) the \hika two-forms are
deleted from the compact support cohomology groups. However the
Hirzebruch signature is what it is, hence also other three selfdual
two-forms have to be deleted as being non square-integrable, in order to
maintain the value of $\tau$.

The ``Hodge diamonds'' for the usual and ${\bf L}^2$
Dolbeaut cohomology groups are respectively given by:
\begin{equation}
\begin{array}{c}
1\\ 0 \hspace{.6cm} 0 \\ 1 \hspace{.6cm} \tau+4 \hspace{.6cm} 1 \\
0 \hspace{.6cm} 0 \\ 0
\end{array}
\hspace{3cm}
\begin{array}{c}
0\\ 0 \hspace{.6cm} 0 \\ 0 \hspace{.6cm} \tau \hspace{.6cm} 0 \\
0 \hspace{.6cm} 0 \\ 1
\end{array}
\label{diamonds}
\end{equation}
Note that, from eq.(\ref{Euchar}), $\chi=\tau +1$.

In the (4,4) SCFT corresponding to a non-compact gravitational instanton
we expect therefore to be able to distinguish four of the
$\Psi_A$ as giving rise to ``non-normalizable'' deformations. We will
see how this is realized in the case of ALE spaces.
\vskip 0.2cm\noindent
\underline{\sl ALE spaces}
\vskip 0.2cm
The most natural gravitational analogues of the Yang-Mills instantons
would be represented by Riemannian manifolds geodesically complete and
such that
\begin{enumerate}
\item the curvature 2-form is (anti)selfdual;
\item the metric approaches the Euclidean metric at infinity; that is,
in polar coordinates $(r,{\bf \Theta})$ on ${\bf R}^4$
\begin{equation}
 g_{\mu\nu}(r,{\bf \Theta})=\delta_{\mu\nu} + O(r^{-4})
\label{asymetric}
\end{equation}
\end{enumerate}
This would agree with the ``intuitive'' picture of instantons as being
localized in finite regions of space-time. The above picture is verified
however only modulo an additional subtlety: the base manifold has a
boundary at infinity $S^3/\Gamma$, $\Gamma$ being a
finite group of identifications. ``Outside the core of the
instanton'' the manifold looks like ${\bf
R}^4/\Gamma$ instead of ${\bf R}^4$ .
This is the reason of the name given to these spaces: the
asymptotic behaviour is only {\it locally} euclidean. The unique {\it globally}
euclidean gravitational instanton is euclidean four-space itself.

This kind of behaviour is easily seen in the simplest of these metrics,
the Eguchi-Hanson metric \cite{eguchihanson}:
\begin{equation}
ds^2={ dr^2\over 1-\left({a\over r}\right)^4} + r^2 (\sigma_x^2
+\sigma_y^2)+r^2\left[1-\left({a\over r}\right)^4\right]\sigma_z^2,
\label{ehmetric1}
\end{equation}
where $a$ is a real constant, $\sigma_x,\sigma_y,\sigma_z$ are
Maurer-Cartan forms of $SU(2)$ realized in terms of the Euler angles
given by the angles of the polar coordinates on ${\bf R}^4$ $\theta,\phi,\psi$.
By changing the radial coordinate: $u^2=r^2\left[1-
\left({a\over r}\right)^4\right]$
the apparent singularity at $r=a$ is moved to $u=0$:
\begin{equation}
ds^2={du^2\over \left[1+\left({a\over r}\right)^4\right]^2}+u^2
\sigma_z^2 + r^2 (\sigma_x^2 + \sigma_y^2)\, ;
\label{ehmetric2}
\end{equation}
Since near $u=0$\hspace{2pt} $ds^2\simeq {1\over 4}du^2+{1\over
4}u^2(d\psi +cos \theta d\phi )^2+{a^2\over 4}(d\theta^2+sin^2\theta
d\phi^2)$, at fixed $\theta,\phi$ the singularity in $ds^2\simeq {1\over
4}(du^2+u^2d\psi^2)$ looks like the removable singularity due to the use
of polar coordinates in ${\bf R}^2$, {\it provided that} $0\leq
\psi<2\pi$, which is {\it not} the range assumed by the polar angle $\psi$
on ${\bf R}^4$; in this case the range is instead $0\leq\psi<4\pi$. Thus
opposite points on the constant-radius slices are to be identified, and
the boundary at infinity is $S^3/{\bf Z}_2$

Subsequent work, leading to the construction of the
``multi-Eguchi-Hanson'' metrics \cite{multieh} and their
reinterpretation in terms of a twistor construction \cite{hitchin},
culminating with the papers by Kronheimer \cite{kro1,kro2},
established the following picture.

Every ALE space is determined by its group of identifications $\Gamma$,
which must be a finite Kleinian subgroup of $SU(2)$.
Kronheimer described indeed
manifolds having such a boundary; he showed that in principle an
unique selfdual metric can be obtained for for each of these manifolds
\cite{kro1} and, moreover, that  every selfdual metric approaching
asymptotically the euclidean one can be recovered in such a manner
\cite{kro2}.
The classification of $SU(2)$ finite subgroups is well-known, and is
reviewed in the following section.

\section{The Kleinian subgroups of $SU(2)$}
\label{grouptheory}

The classification of the finite (Kleinian) subgroups of $SU(2)$ is a
classical result of nineteenth century mathematics \cite{KleinIcosa}
and can be related in a one-to-one fashion to the
twentieth century classification of simply laced Lie algebras
(ADE classification), as much as to the two thousand years  old Platonic
classification of regular polygons, dihedra and polyhedra
\cite{PlatoTimaeus}. The explicit construction of the ALE manifolds as
HyperK\"ahler quotients
obtained in 1989 by Kronheimer \cite{kro1,kro2} relies heavily on the
algebraic structure of the
Kleinian groups and puts into evidence the crucial
identification between the most relevant
topological  number of the manifold, namely its Hirzebruch
signature $\tau$  and the number $r$ of conjugacy classes
of the finite group $\Gamma$.
Furthermore it is this identification that provides a clue
for the construction of the corresponding
(4,4) conformal field theories.
Choosing complex coordinates $z_1=x-iy,z_2=t+iz$ on
${\bf R}^4\sim{\bf C}^2$, and representing a point $(z_1,z_2)$ by a
quaternion (see sec. 4), the group $SO(4)\sim SU(2)_L\times SU(2)_R$, which is
the isometry group of the sphere at infinity, acts on the
quaternion by matrix multiplication:
\begin{equation}
\twomat{z_1}{i\bar z_2}{iz_2}{\bar z_1}
\hspace{3pt}\longrightarrow\hspace{3pt} M_1 \cdot
\twomat{z_1}{i\bar z_2}{iz_2}{\bar z_1}
\cdot M_2
\label{matrixaction}
\end{equation}
the element $M\in SO(4)$ being represented as $(M_1\in SU(2)_L, M_2\in
SU(2)_R)$.
The group $\Gamma$ can be seen as a finite subgroup of $SU(2)_L$,
acting on ${\bf C}^2$ in the natural way by its two-dimensional
representation:
\begin{equation}
\forall U\in\Gamma\subset SU(2),
\hspace{15pt} U:\hspace{3pt}
{\bf v}=\twovec{z_1}{z_2}
\hspace{3pt} \longrightarrow
U{\bf v} =\twomat {\alpha}{i \beta}{i{\bar \beta}}{\bar \alpha}
\twovec {z_1} {z_2}\, .
\label{grouptheory3}
\end{equation}
In characterizing such finite subgroups it appears a
Diophantine equation which is just the same one which is encountered in
the classification of the possible simply laced Dynkin diagrams \cite{cristal}.
As a result, the possible finite subgroups of $SU(2)$ are organized in two
infinite series and three exceptional cases; each subgroup $\Gamma$ is in
correspondence with a simply laced Lie algebra $\cG$, and we write
$\Gamma(\cG)$ for it. One series is given by the
cyclic subgroups groups of order $k+1$, related to $A_k$; the other
series is that of the dihedral subgroups containing a cyclic subgroup of
order $k$, related to $D_{k+2}$; the remaining three subgroups $\Gamma (E_6)
\, \approx \, {\cal T}$,
$\Gamma(E_7) \, \approx \, {\cal O}$ and $\Gamma(E_8) \, \approx \,
{\cal I}$ have order $12$, $24$ and $60$, respectively, and they correspond
to the binary extensions of the
Tetrahedron, Octahedron and Icosahedron symmetry groups.
More specifically, the ADE classification of $SU(2)$ finite subgroups is
obtained by considering the ``poles'' $P_i$ identified by the subgroup
$\Gamma$, where $P_i=\{\lambda v_i\}$ $i=1,2$ is, by definition, the
one-dimensional subspace of ${\bf C}^2$ identified by an eigenvector
$v_i$ of some element $\gamma\in\Gamma$. Since each nontrivial element of the
group $SU(2)$ has two distinct orthogonal eigenvectors, a finite subgroup
$\Gamma$ of order $n$ singles out $2n-2$
poles that are not necessarily distinct. (For instance, in the case of the
${\bf Z}_n$ subgroup generated by the rotation of an angle ${2\pi\over n}$
around some axis, all elements of the group single out the same pair of
distinguished poles). If we call equivalent two poles $P_i$ and $P_j$ that
are mapped one into the other by some elements $\gamma\in\Gamma$
($P_i=\gamma P_j$), then we can distribute the $2n-2$ poles into
$r$ equivalence classes ${\cal C}_\alpha$ ($\alpha=1,\ldots r$). We name
$m_\alpha$ the number of distinguished poles that belong to the class
${\cal C}_\alpha$ and $q_\alpha$ the total number of poles in ${\cal
C}_\alpha$,
counting as different also coincident poles. A simple argument for which
we refer to \cite{cristal} shows that $q_\alpha=m_\alpha(k_\alpha-1)$, where
$k_\alpha$ is the order of the stability subgroup $K_P$ of any pole $P$
in ${\cal C}_\alpha$. Indeed, one sees that all equivalent poles have
conjugate stability subgroups, so that their order depends only on the
equivalence class. Furthermore, the stability subgroups $K_P$ are cyclic groups
$K_P\approx {\bf Z}_{k_\alpha}$.
Using this information, one immediately obtains
\begin{equation}
2n-2=\sum_{\alpha=1}^r m_\alpha (k_\alpha -1).
\label{equazione}
\end{equation}
Another simple argument shows that for each equivalence class
${\cal C}_\alpha$ we have $m_\alpha k_\alpha=n$.
Indeed, it is sufficient to decompose
the group $\Gamma=K_P+g_1 K_P+\ldots+G_s K_P$ into
cosets with respect to the stability subgroup $K_P$ of any pole $P$
in ${\cal C}_\alpha$, and to remark that the number $s$ of cosets is
just equal to the number $m_\alpha$ of
distinct poles equivalent to $P$. (By definition, $g_i P$ is different
from $P$ and equivalent to $P$). Dividing eq.(\ref{equazione})
by $n$ and using this information, we obtain the Diophantine equation leading
to the ADE classification, namely
\begin{equation}
2 \left ( \, 1 \, - \, \o{1}{n} \, \right ) ~=~\sum_{\alpha=1}^{q}\,
 \left ( 1 \, - \, \o{1}{k_\alpha} \, \right )\, .
\label{grouptheory5}
\end{equation}
Since by definition $k_\alpha \ge 2$, a little bit of elaboration of
eq. (\ref{grouptheory5})
shows that it implies, as only possible cases $q=2$ or $q=3$. In the case $q=2$
 eq. (\ref{grouptheory5}) reduces to
\begin{equation}
\o{1}{k_1} \, + \, \o{1}{k_2} \, = \, \o{2}{n}
\label{grouptheory6}
\end{equation}
whose only solution is $k_1J\, = \, k_2 \, = \, k+1$.
This is the $A_{k}$ solution and the
corresponding subgroup $\Gamma ( A_{k})\, \approx \,
{\bf Z}_{k+1}$ is cyclic of order $k+1$,
 its elements being the matrices given in eq.(\ref{grouptheory7})
Indeed the two equivalence classes of poles are spanned by
the two eigenvectors of
the single ${\bf Z}_{k+1}$ generator that are common to all the group elements.
\par
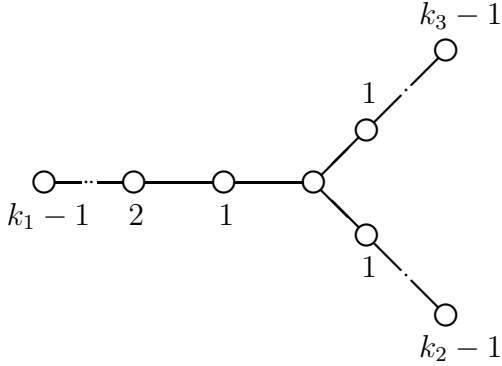
\begin{figure}
\label{dotfigure}
\begin{picture}(200,100)(-170,-66)
\thicklines
\multiput(0,0)(-34,0){4}{\circle{8}}
\multiput(20,20)(0,-40){2}{\circle{8}}
\multiput(50,50)(0,-100){2}{\circle{8}}
\multiput(-4,0)(-34,0){2}{\line(-1,0){26}}
\put(-72,0){\line(-1,0){10}}
\put(-98,0){\line(1,0){10}}
\multiput(-84,0)(-2,0){2}{\circle*{1}}
\put(3,3){\line(1,1){14}}
\put(3,-3){\line(1,-1){14}}
\put(23,23){\line(1,1){10}}
\put(23,-23){\line(1,-1){10}}
\put(47,47){\line(-1,-1){10}}
\put(47,-47){\line(-1,1){10}}
\put(35,35){\circle*{1}}
\put(35,-35){\circle*{1}}
\put(-36,-16){$1$}
\put(-70,-16){$2$}
\put(-116,-16){$k_1-1$}
\put(18,30){$1$}
\put(18,-36){$1$}
\put(40,60){$k_3-1$}
\put(40,-66){$k_2-1$}
\end{picture}
\caption{\sl Interpretation of the numbers $k_i$}
\end{figure}

In the $q=3$ case the Diophantine equation (\ref{grouptheory5}) becomes:
\begin{equation}
\o{1}{k_1} \, + \, \o{1}{k_2} \, + \, \o{1}{k_3} \, = \, 1 \, + \, \o{2}{n}
\label{grouptheory8}
\end{equation}
admitting the infinite $D_{k+2}$ solution $k_1 = k ,\, k_2 = k_3 = 2$
and the three exceptional $E_6$, $E_7$ and $E_8$ solutions respectively
given by $k_1=3 , \, k_2= 3, \, k_3=2$
or $k_1=4, \, k_2= 3, \, k_3=2$, or $k_1=5 , \, k_2= 3, \, k_3=2$.
The reason why these solutions are associated with the names of
the simply laced Lie algebras is that the numbers
$k_i - 1$ can be interpreted as the number of dots in each
simple chain departing from the
dot at the center of a node in a corresponding Dynkin diagram (see
\ref{dotfigure}).

The group $\Gamma (D_{k+2})$ is the dihedral subgroup. Its order is
\begin{equation}
| \, \Gamma (D_{k+2}) \, | ~=~4 \, k
\label{grouptheory9}
\end{equation}
and it contains a cyclic subgroup of order $k$ that we name $K$.
Its index in $\Gamma (D_{k+2})$
is  two. The elements of $\Gamma (D_{k+2})$ that are not in $K$
are of period equal to two since
$k_2 \, = \, k_3 \, = 2$. Altogether the elements of the dihedral group are
the  matrices given in eq.(\ref{grouptheory10}).
The remaining three subgroups $\Gamma (E_6) \, \approx \, {\cal T}$,
$\Gamma(E_7) \, \approx \, {\cal O}$ and $\Gamma(E_8) \, \approx \,
{\cal I}$ can be similarly described.

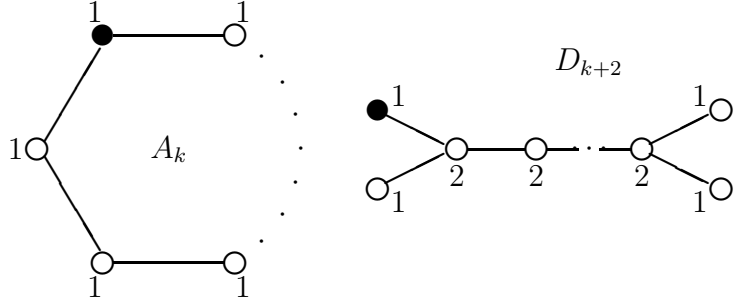
\begin{figure}[tb]
\label{dynfigure1}
\begin{center}
\begin{picture}(280,110)(-60,-60)
\thicklines
\put(25,43){\circle{8}}
\put(-25,43){\circle*{8}}
\put(-50,0){\circle{8}}
\put(-25,-43){\circle{8}}
\put(25,-43){\circle{8}}
\put(35,35){\makebox(0,0){$\cdot$}}
\put(43,25){\makebox(0,0){$\cdot$}}
\put(48,13){\makebox(0,0){$\cdot$}}
\put(50,0){\makebox(0,0){$\cdot$}}
\put(48,-13){\makebox(0,0){$\cdot$}}
\put(43,-25){\makebox(0,0){$\cdot$}}
\put(35,-35){\makebox(0,0){$\cdot$}}
\put(-21,43){\line(1,0){42}}
\put(-47,3){\line(3,5){21}}
\put(-47,-3){\line(3,-5){21}}
\put(-21,-43){\line(1,0){42}}
\put(28,52){\makebox(0,0){1}}
\put(-28,52){\makebox(0,0){1}}
\put(-58,0){\makebox(0,0){1}}
\put(-28,-52){\makebox(0,0){1}}
\put(28,-52){\makebox(0,0){1}}
\put(0,0){\makebox(0,0){$A_{k}$}}
\put(75,-30){
\begin{picture}(150,50)(0,-30)
\thicklines
\put(67,30){$D_{k+2}$}
\multiput(30,0)(30,0){2}{\circle{8}}
\put(30,-10){\makebox(0,0){2}}
\put(60,-10){\makebox(0,0){2}}
\put(34,0){\line(1,0){22}}
\put(64,0){\line(1,0){12}}
\put(80,0){\makebox(0,0){$\cdots$}}
\put(84,0){\line(1,0){12}}
\put(100,0){\circle{8}}
\put(100,-10){\makebox(0,0){2}}
\put(0,15){\circle*{8}}
\put(8,20){\makebox(0,0){1}}
\put(0,-15){\circle{8}}
\put(8,-20){\makebox(0,0){1}}
\multiput(130,15)(0,-30){2}{\circle{8}}
\put(122,20){\makebox(0,0){1}}
\put(122,-20){\makebox(0,0){1}}
\put(3,13){\line(2,-1){22}}
\put(3,-13){\line(2,1){22}}
\put(103,2){\line(2,1){22}}
\put(103,-2){\line(2,-1){22}}
\end{picture}}
\end{picture}
\caption{\sl Extended Dynkin diagrams of the infinite series}
\end{center}
\end{figure}

\begin{figure}[tb]
\label{dynfigure2}
\begin{center}
\begin{picture}(300,135)(0,-10)
\thicklines
\put(5,25){$E_8\leftrightarrow {\cal I}$}
\multiput(0,0)(30,0){7}{\circle{8}}
\put(210,0){\circle*{8}}
\put(0,-10){\makebox(0,0){2}}
\put(30,-10){\makebox(0,0){4}}
\put(60,-10){\makebox(0,0){6}}
\put(90,-10){\makebox(0,0){5}}
\put(120,-10){\makebox(0,0){4}}
\put(150,-10){\makebox(0,0){3}}
\put(180,-10){\makebox(0,0){2}}
\put(210,-10){\makebox(0,0){1}}
\multiput(4,0)(30,0){7}{\line(1,0){22}}
\put(60,4){\line(0,1){22}}
\put(60,30){\circle{8}}
\put(52,30){\makebox(0,0){3}}

\put(140,45){
\begin{picture}(180,50)(0,0)
\thicklines
\put(5,25){$E_7\leftrightarrow {\cal O}$}
\multiput(0,0)(30,0){6}{\circle{8}}
\put(180,0){\circle*{8}}
\put(0,-10){\makebox(0,0){1}}
\put(30,-10){\makebox(0,0){2}}
\put(60,-10){\makebox(0,0){3}}
\put(90,-10){\makebox(0,0){4}}
\put(120,-10){\makebox(0,0){3}}
\put(150,-10){\makebox(0,0){2}}
\put(180,-10){\makebox(0,0){1}}
\multiput(4,0)(30,0){6}{\line(1,0){22}}
\put(90,4){\line(0,1){22}}
\put(90,30){\circle{8}}
\put(82,30){\makebox(0,0){2}}
\end{picture}}

\put(-20,80){
\begin{picture}(120,50)(0,15)
\thicklines
\put(5,25){$E_6\leftrightarrow {\cal T}$}
\multiput(0,0)(30,0){5}{\circle{8}}
\put(0,-10){\makebox(0,0){1}}
\put(30,-10){\makebox(0,0){2}}
\put(60,-10){\makebox(0,0){3}}
\put(90,-10){\makebox(0,0){2}}
\put(120,-10){\makebox(0,0){1}}
\multiput(4,0)(30,0){4}{\line(1,0){22}}
\put(60,4){\line(0,1){22}}
\put(60,30){\circle{8}}
\put(52,30){\makebox(0,0){2}}
\put(52,60){\makebox(0,0){1}}
\put(60,34){\line(0,1){22}}
\put(60,60){\circle*{8}}
\end{picture}}
\end{picture}
\caption{\sl Exceptional extended Dynkin diagrams}
\end{center}
\end{figure}
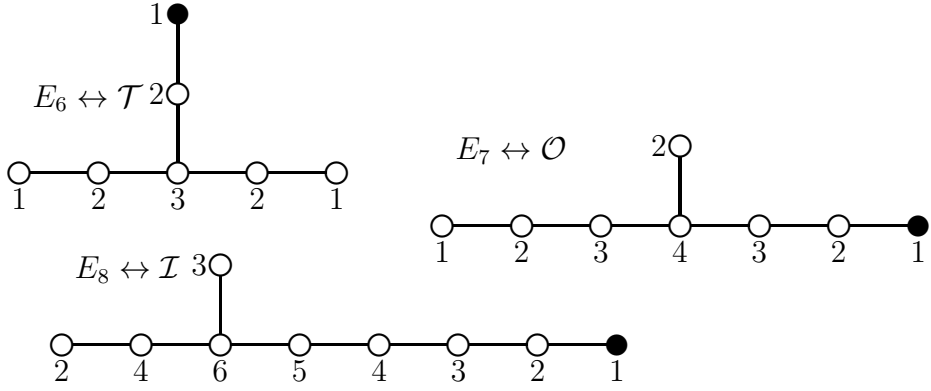

Let us now consider more closely the algebraic structure of these groups,
in particular in relation with their conjugacy classes, whose number equals
the Hirzebruch signature of the corresponding ALE manifold. We begin by fixing
notations. For any finite group $\Gamma$ we denote by $K_i$,  $(i=1,....,r)$
the conjugacy classes of its elements: ${\rm iff}~
\gamma_1 \, \gamma_2 \, \in \, K_i \, {\rm then} ~\exists \, h \, \in \,
\Gamma \, / \,\gamma_1 \, = \, h^{-1} \, \gamma_2 \, h$ and  we name $g_i \,
= \, | K_i | $ the order of the
$i$-th conjugacy class. One obviously has $| \Gamma |J\,=
\, g \, = \, \sum_{i=1}^{r} \, g_i$. For any representation $D$ of $\Gamma$,
we denote by $\left \{ \chi^{(D)}_1 \, , \, ...... \, , \, \chi^{(D)}_r \,
\right \}$ its character vector
where $\chi^{(D)}_i \, = \, \tr \, \left ( \, D(\gamma_i)J\, \right )$ is the
trace of any representative of the $i$-th class. As it is well-known the
number of conjugacy classes $r$ equals the number of irreducible
representations: we name these latter  $D_{\mu}$
($\mu \, = \, 1,....,\, r$). The square matrix $\chi^{\mu}_i$ whose rows
are the character vectors of the irreducible representations is named the
{\it character table}. It satisfies the orthogonality relations
\begin{equation}
\sum_{\mu =1}^{r} \, \chi_{i}^{\mu} \, \chi^{\mu}_{j} = \o{g}{g_i} \,
\delta_{ij}\nonumber\hskip 2pt , \hskip 12pt
\sum_{i =1}^{r} \, \chi_{i}^{\mu} \, \chi^{\nu}_{i} = g \, \delta^{\mu\nu}
\label{grouptheory13}
\end{equation}
that imply the following sum rule for the dimensions $n_\mu =\tr D_\mu (\un)$
of the irreducible representations:
\begin{equation}
\sum_{\mu =1}^{r} \, n_\mu^2 \, = \, g \, = \, | \Gamma | \, .
\label{grouptheory14}
\end{equation}
Relevant to our use of the Kleinian groups are also the $g$-dimensional
regular representation $R$, whose basis vectors $e_\gamma$ are in one-to-one
correspondence with the group elements $\gamma$ and transform as
\begin{equation}
 R(\gamma) \, e_\delta ~=~e_{\gamma \cdot \delta} ~~~~~\forall \,
\gamma \, , \, \delta \, \in \, \Gamma
 \label{grouptheory15}
 \end{equation}
and the  2-dimensional defining representation ${\cal Q}$ which is  obtained
by regarding the group $\Gamma$ as an $SU(2)$ subgroup [that is, $\cQ$
is the representation which acts in eq.(\ref{grouptheory3})].
The character table allows to reconstruct the decomposition
coefficients of any representation
along the irreducible representations.
If $D=\bigoplus_{\mu =1}^{r} \, a_\mu \, D_\mu$
we have $a_\mu= \o{1}{g} \, \sum_{i =1}^{r} \, g_i \, \chi^{(D)}_{i} \,
\chi^{(\mu) \, \star}_{i}$. For the Kleinian groups $\Gamma$ a
particularly important case is
the decomposition of the tensor product of an irreducible
representation $D_\mu$ with the
defining 2-dimensional representation ${\cal Q}$.  It is indeed at the
level of this decomposition that the relation between these groups
and the simply laced Dynkin diagrams is more explicit \cite{mckay}.
Furthermore this decomposition plays a
crucial role in the explicit construction of the ALE manifolds
\cite{kro1}. Setting
\begin{equation}
{\cal Q} \, \otimes \, D_\mu ~=~\bigoplus_{\nu =0}^{r} \, A_{\mu \nu} \, D_\nu
\label{grouptheory16}
\end{equation}
where $D_0$ denotes the identity representation,
one finds that the matrix ${\bar c}_{\mu\nu}=2\delta_{\mu\nu}-A_{\mu\nu}$
is the {\it extended Cartan matrix} relative to {\it  extended
Dynkin diagram} corresponding
to the given group. We remind the reader the the extended Dynkin diagram of
any simply laced Lie algebra is obtained by adding to the {\it  dots}
representing  the {\it  simple roots}
$\left \{ \, \alpha_1 \, ......\, \alpha_r \,  \right \}$
an {\it  additional dot} (marked black in Fig.s \ref{dynfigure1},
\ref{dynfigure2}) representing the negative of
the highest root $\alpha_0 \, = \, \sum_{i=1}^{r} \, n_i \, \alpha_i$
($n_i$ are the Coxeter
numbers). We see thus an correspondence between the non-trivial conjugacy
classes $K_i$ or equivalently the non-trivial irrepses of the group
$\Gamma(\cG)$ and the simple roots of $\cG$. In this correspondence,
as we have already remarked the extended
Cartan matrix provides us with the Clebsch-Gordan coefficients
(\ref{grouptheory16}), while
the Coxeter numbers $n_i$ express the dimensions of the
irreducible representations. All
these informations are summarized in Fig.s 2,3 where the
numbers $n_i$ are attachedto each of the dots: the number $1$
is attached to the extra dot because it stands for the
identity representation.

\par
Let us now briefly consider the structure of the irreducible
representations and of the character tables.
\par\noindent
\underline{\sl $A_{k}$-series}\hskip 0.2cm
In this case the defining 2-dimensional representation ${\cal Q}$ is given
by  the matrices
\begin{equation}
\gamma_l \, \in \, \Gamma (A_{k})~~~~; ~~~~\gamma_l \, = \,
{\cal Q}_l \, \eqdef \,\twomat {e^{2\pi i l/(k+1)}}{0}{0}
{e^{-\, 2\pi i l/(k+1}} \,\{ l=1,.....,k\}\, .
\label{grouptheory7}
\end{equation}
It is not irreducible since all irreducible representations
are one-dimensional as one sees from Fig.\ref{dynfigure1}.
In the $j$-th irreducible representation the $1\times 1$-matrix
representing the $l$-th element of the group is
\begin{equation}
D^{(j)}\left( e_l \right) ~=~\nu^{jl} ~~~;
{}~~~{\rm where} ~~\nu={\rm exp}{ {\o{2\pi i}{k}} }\, .
\label{grouptheory17}
\end{equation}
The $(k+1)\times (k+1)$ array of phases $\nu^{jl}$ appearing in
the above equation is the character table.
Given the ${\bf C}^2$ carrier space of the defining representation
[see eq.s (\ref{grouptheory3})] it is fairly easy
to construct three algebraic invariants, namely
\begin{eqnarray}
z&=&z_1 \, z_2\nonumber\\
x&=&\left ( z_1 \right )^{k+1}\nonumber\\
y&=&\left ( z_2 \right )^{k+1}
\label{grouptheory19}
\end{eqnarray}
that satisfy the polynomial relation
\begin{equation}
W_{A_k} \, \left ( \,  x, \, y, \, z \right ) \, \eqdef \,
x \, y \, - z^{k+1} \, = \, 0 \, .
\label{grouptheory20}
\end{equation}
\par
As stressed in the introduction the polynomial constraint
$W_{\Gamma} (x,y,z) \, = \, 0$ plays an important role in the construction
of the ALE manifolds and of the associated (4,4)-conformal field-theory.
Indeed, as we are going to see in the next sections, the vanishing
locus in ${\bf C}^3$ of the {\it potential} $W_{\Gamma}(x,y,z)$ coincides
with the space of equivalence classes ${\bf C}^2 / \Gamma$,
that is with the singular orbifold
limit of the self-dual manifold ${\cal M}_{\Gamma}$. According to the
standard procedure of deforming singularities
\cite{singularityliterature} there is a corresponding family of
smooth manifolds ${\cal M}_{\Gamma}\left(t_1, t_2 , ....., t_r \right)$
obtained as the vanishing locus $Z_0\in {\bf C}^3$ of a
{\it deformed  potential}:
\begin{equation}
{\tilde W}_{\Gamma} \left (  x,y,z ; \, t_1, t_2 , .....,
t_r \right )~=~W_{\Gamma}(x,y,z)\, + \,\sum_{i=1}^{r} \ t_i \,
{\cal P}^{(i)}(x,y,z)
\label{grouptheory21}
\end{equation}
where $t_i$ are complex numbers (the moduli of the complex
structure of ${\cal M}_{\Gamma}$)
and ${\cal P}^{(i)}(x,y,z)$ is a basis spanning the chiral ring
\begin{equation}
{\cal R}=\o{{\bf C}[x,y,z]}{\partial W}
\label{grouptheory22}
\end{equation}
of polynomials in $x,y,z$ that do not vanish upon use of the
vanishing relations $\partial_x \, W
\,= \, \partial_y \, W \, = \, \partial_z  \, W \, = \, 0$. It is a
matter of fact that the dimension of this chiral ring $| {\cal R} |$ is
precisely equal to the number of non-trivial conjugacy classes
(or of non trivial irreducible representations)  of the finite group $\Gamma$.
{}From the geometrical point of view this implies an
identification between the number of complex
structure deformations of the ALE manifold and the number $r$ of
non-trivial conjugacy classes discussed above. From the (4,4) CFT
viewpoint this relation implies that $r$ must also be the number
of short representations, whose last components (moduli operators)
can be used to deform the theory. In other words we have $\tau =r$,
where $\tau$ is the Hirzebruch signature.
Indeed in the language of algebraic geometry the singular orbifold $\csg$
corresponding to the vanishing locus $Z_0$ of the potential $W$
admits an equivariant minimal resolutions of singularity
$Z\stackrel{\lambda}{\longrightarrow}Z_0$, where $Z$ is a smooth variety,
$\lambda$ is an isomorphism outside the singular point $\{0\}\in
Z_0$ and it is a proper map  such that $\lambda^{-1}(Z_0-0)$ is
dense in $Z$. The fundamental fact is that the exceptional divisor
$\lambda^{-1}(0)\subset
Z$ consists of a set of irreducible curves $c_{\alpha}, \alpha=1,\ldots
r$ which can be put in correspondence with the vertices of the Dynkin
diagram (the non-extended one) of the simple Lie Algebra corresponding
to $\Gamma$ as above. Each $c_{\alpha}$ is isomorphic
to a copy of $\c{\bf P}^1$; the intersection matrix of these non-trivial
two-cycles is the negative of the Cartan matrix:
\begin{equation} c_{\alpha}\cdot c_{\beta}=\bar c_{\alpha \beta}\, .
\label{cartan}\end{equation}
Kronheimer construction, described in section \ref{construction}, shows that
the base manifold $\cM$ of an
ALE space is diffeomorphic to the space $Z$ supporting the resolution of
the orbifold $Z_0\sim\csg$, see section \ref{construction}. Therefore
the equation (\ref{cartan}) applies to the generators of the second homology
group of $\cM$. In particular we see that
\begin{eqnarray}
\tau & = &dim H^2_{\bf c}(X)=dim H_2(X)=\nonumber\\
     & = &{\rm rank\,\, of\,\, the\,\, corresponding\,\, Lie\,\,
          Algebra}=\nonumber\\
     & = &\#\,\,{\rm non\,\, trivial\,\, conj.\,\, classes\,\,
          in}\,\,\Gamma= |{\cal R}|\label{tau} \, .
\end{eqnarray}

The chiral ring of the potential (\ref{grouptheory30}) has $k+2$
elements just matching the number of non-trivial conjugacy classes.
According to our previous discussion $k+2$
will also be the number of short representations in the
corresponding (4,4) conformal field-theory.
\par
These results are summarized in Table \ref{kleinale} which compares
the algebraic information on the Kleinian group structure with the
classical results on the topology of the ALE
manifolds obtained in the late seventies by means of the index theorems
(see \cite{ehphysrep}). Indeed in  the last columns of
Table \ref{kleinale} 2 we list the Hirzebruch signature $\tau$ ,
the Euler character $\chi$ and the spin 3/2 index $I_{3/2}$.
As one sees, the Hirzebruch signature is always equal to the
dimension of the chiral ring, which also equals the number of
conjugacy classes of the Kleinian group. The spin 3/2
index counts the normalizable gravitino zero-modes and turns
out to be equal to $2 \tau\,
= \, |{\cal R}|$. This is in agreement with the results of \cite{previous}.
\par\noindent
\underline{\sl $D_{k+2}$-series} \hskip 0.3cm
Abstractly the binary extension $D_{k+2}$ of the dihedral group
could be described introducing
the generators ${\cal A}, \, {\cal B}, \, {\cal Z}$ and setting the relations:
\begin{eqnarray}
{\cal A}^k&=&{\cal B}^2~=~\left ( {\cal A} \, {\cal B}
\right )^2~=~{\cal Z}\nonumber\\{\cal Z}^2&=&\un \, .
\label{grouptheory23}
\end{eqnarray}
The $4k$ elements of the group are given by the following matrices:
\begin{eqnarray}
F_l&=&\twomat {e^{i l \pi/k}}{0}{0}{e^{ - i l \pi/k}}~~~~;
{}~~~(l=0,1,2,.....,2k-1)\nonumber\\
G_l&=&\twomat{0}{ i \, e^{-i l \pi/k}}{i\, e^{ i l \pi/k}}{0}~~~~;
{}~~~(l=0,1,2,.....,2k-1)
\label{grouptheory10}
\end{eqnarray}
In terms of them the generators are identified as follows:
\begin{equation}
F_0~=~\un~~~;~~~F_1~=~{\cal A}~~~;~~~F_k~=~{\cal Z}~~~;~~~G_0~=~{\cal
B}\,\,.
\label{grouptheory24}
\end{equation}
There are exactly $r~=~k+3$ conjugacy classes
\begin{enumerate}
\item $K_e$ contains only the identity $F_0$
\item $K_Z$ contains the central extension ${\cal Z}$
\item $K_{G \, even}$ contains the elements $G_{2\nu}$ ($\nu=1,....,k-1$)
\item $K_{G \, odd}$ contains the elements $G_{2\nu +1}$ ($\nu=1,....,k-1$)
\item  the $k-1$ classes $K_{F_\mu}$: each of these classes
contains the pair of elements $F_\mu$ and
$F_{2k - \mu}$ for ($\mu=1,....,k-1$).
\end{enumerate}
Correspondingly the $D_{k+2}$ group admits $k+3$ irreducible
representations $4$ of which are
1-dimensional while $k-1$ are 2-dimensional. We name them as follows:
\begin{equation}
\cases{D_e ~;~D_Z~;~D_{G \, even}~;~D_{G \, odd}~~~~;~~~~1-dimensional\cr
D_{F_1}~;~.........~;~D_{F_{k-1}}~~~~;~~~~~2-dimensional \,\, .\cr}
\label{grouptheory26}
\end{equation}
The combinations of the ${\bf C}^2$ vector components $(z_1, z_2)$
that transform in
the four 1-dimensional representations are easily listed:
\begin{eqnarray}
D_e&\longrightarrow&~~~ |z_1|^2 + |z_2|^2\nonumber\\
D_Z&\longrightarrow&~~~z_1 \, z_2\nonumber\\
D_{G \, even}&\longrightarrow&~~~z_1^k +z_2^k\nonumber\\
D_{G \, odd}&\longrightarrow&~~~z_1^k -z_2^k\, .
\label{grouptheory27}
\end{eqnarray}
The matrices of the $k-1$ two-dimensional representations  are obtained
in the following way. In the $DF_s$ representation, $s=1,\ldots k-1$,
the generator $\cal A$, namely the group element $F_1$, is represented
by the matrix $F_s$. The generator $\cal B$ is instead represented by
$(i)^{s-1}G_0$ and the generator $\cZ$ is given by $F_{sk}$, so that:
\begin{eqnarray}
DF_s\,(F_j)&=&F_{sj}\nonumber\\
DF_s\,(G_j)&=&(i)^{s-1}G_{sj}\, .
\label{grouptheory28}
\end{eqnarray}
The character table is immediately obtained and it is displayed in
Table \ref{dktable}.
\begin{table}\caption{\sl Character table of the Group $D_{k+2}$}
\label{dktable}
\begin{center}
\begin{tabular}{||c||c|c|c|c|c|c|c||}\hline
$~$. & $KE$ &$KZ$& $KG_{e}$&$KG_{o}$ & $KF_1$
&$\cdots$ &$KF_{k-1}$ \\
\hline
\hline
$DE$  &  $ 1$ & $   1$     &  $    1$ &  $     1$  & $  1$
& $\cdots$ &        $  1$\\
\hline
$DZ$  &  $ 1$ & $   1$     &  $ -1$  & $    -1$  & $  1$
&  $\cdots$&         $ 1$   \\
\hline
$DG_{e}$ & $ 1$  & $ (-1)^k$ & $ i^k$  & $ -i^k$   & $ (-1)^1$
&$\cdots$&$(-1)^{k-1}$   \\
\hline
$DG_{o}$ & $ 1$  & $ (-1)^k$ & $ -i^k$  & $ i^k$   & $ (-1)^1$
&$\cdots$ &$(-1)^{k-1}$   \\
\hline
$~$ & $ ~$  & $ ~$ & $ ~$  & $ ~$   & $ ~$
&$~$&$~$   \\
$DF_1$ & $ 2$  & $ (-2)^1$ & $ 0$  & $ 0$   & $ 2\, Cos\o{\pi}{k}$
&$\cdots$&$2\, Cos \o{(k-1) \pi}{k}$   \\
$~$ & $ ~$  & $ ~$ & $ ~$  & $ ~$   & $ ~$
&$~$&$~$   \\
\hline
$\vdots$ & $\vdots$  & $\vdots$ & $\vdots$  & $\vdots$   & $\vdots$
&$\ddots$
&$\vdots$   \\
\hline
$~$ & $ ~$  & $ ~$ & $ ~$  & $ ~$   & $ ~$
&$~$&$~$   \\
$DF_{k-1}$ & $ 2$  & $ (-2)^{k-1}$ & $ 0$  & $ 0$   & $ 2\, Cos\o{(k-1)\pi}{k}$
&$\cdots$&$2\, Cos \o{(k-1)^2 \pi}{k}$   \\
$~$ & $ ~$  & $ ~$ & $ ~$  & $ ~$   & $ ~$
&$~$&$~$   \\
\hline
\end{tabular}\end{center}
\end{table}
Using the one-dimensional representations (\ref{grouptheory27})
we can define the following invariants:
\begin{eqnarray}
z&=&-(z_1 z_2)^2 \nonumber\\
x&=&\o i2 z_1 z_2 \left(z_1^{2k}-(-1)^kz_2^{2k}\right)\nonumber\\
y&=&\o i2\left(z_1^{2k}+(-1)^kz_2^{2k}\right)
\label{grouptheory29}
\end{eqnarray}
that, in the $D_{k+2}$ case, fulfill the relation
\begin{equation}
W_{D_{k+2}} \, \left ( \,  x, \, y, \, z \right ) \, \eqdef \, x^2 \,  +
\,  y^2 z \, + \, z^{k+1} \, = \, 0\, ,
\label{grouptheory30}
\end{equation}
the analogue of the relation (\ref{grouptheory20}) obtained in the
$A_k$ case. The
chiral ring of the potential (\ref{grouptheory30}) has $k+2$
elements just matching the number of non-trivial conjugacy classes.
According to our previous discussion $k+2$
will also be the number of short representations in the corresponding
(4,4) conformal field-theory.
\par
In a similar way one can retrieve the structure of the irreducible
representations and the potential also for the three exceptional
groups ${\cal T}$, ${\cal O}$ and ${\cal I}$.

\begin{table}\caption{KLEINIAN GROUP versus ALE MANIFOLD properties }
\label{kleinale}
\begin{center}
\begin{tabular}{||l|c|c|c|c|c|r||}\hline
$\Gamma$. & $W(x,y,z)$ & ${\cal R}=\o{{\bf C}[x,y,z]}{\partial W}$
&$|{\cal R}|$ & ${\#} c.~c. $& $\tau\equiv\chi -1$ &$I_{3/2}$  \\ \hline
\hline
$A_k$&$xy - z^{k+1}$&$\{ 1, z,.. $&$k$&$k+1$&$k$&$2k$\\
$~$&$~$&  $.., z^{k-1} \}$&$~$&$~$&$~$&$~$\\ \hline
$D_{k+2}$&$x^2 +y^2 z + z^{k+1}$&$\{ 1, y, z,y^2,$&$k+2$&$k+3$&$k+2$&$2k$\\
$~$&$~$&  $ z^2, ..., z^{k-1} \}$&$~$&$~$&$~$&$+4$\\ \hline
$E_6=$&$x^2+y^3 +z^{4}$&$\{ 1, y,  z,$&$6$&$7$&$6$&$12$\\
${\cal T}$&$~$&  $yz, z^2,yz^{2} \}$&$~$&$~$&$~$&$~$\\ \hline
$E_7=$&$x^2+y^3 +yz^{3}$&$\{ 1, y, z,y^2,$&$7$&$8$&$7$&$14$\\
${\cal O}$&$~$&  $z^2,yz,y^2z \}$&$~$&$~$&$~$&$~$\\ \hline
$E_8=$&$x^2+y^3 + z^{5}$&$\{ 1,y, z,z^2,yz,$&$8$&$9$&$8$&$16$\\
${\cal I}$&$~$&  $z^3,yz^2,yz^3  \}$&$~$&$~$&$~$&$~$\\ \hline
\end{tabular}\end{center}
\end{table}
\vskip 0.2cm
\underline{\sl Number of parameters in the ALE metrics}
\vskip 0.2cm
The number of parameters in a general metric $g_{\mu\nu}$, a part from
the breathing mode, equals the number of zero modes of the Lichnerowitz
operator. These modes modes are
represented by symmetric, traceless, harmonic tensors $\delta g_{\mu\nu}$
that in four dimensions can be obtained as
\begin{equation}\delta g_{\mu\nu}=s_{\mu}^{\hspace{2pt}\rho}a_{\rho\nu}\,
, \label{deltagimunu}\end{equation}
$s_{\mu\nu},a_{\mu\nu}$ being the components of a selfdual (resp.
antiselfdual) harmonic two-form. For \hika four-manifolds, the number
of such modes is clearly
\begin{equation}\#{\rm\,\,traceless \,\, defs.}=b^{2(+)} b^{2(-)}=3\,
b^{2(-)}\, .\label{metricdef1}\end{equation}
A deformation $\delta g_{\mu\nu}$ is called normalizable (${\bf
L}^2$-integrable) if
\begin{equation}
\int_{\cM}\,g^{\mu\nu}g^{\rho\sigma}\delta g_{\mu\rho}\delta g_{\nu\sigma}
\le \infty\, .
\label{normalizability}
\end{equation}
The above deformations, eq.(\ref{deltagimunu}), are not
normalizable when they decrease at radial infinity less strongly
than $r^{-4}$. Adding such a
$\delta g_{\mu\nu}$ to $g_{\mu\nu}$ would destroy the asymptotic
behaviour (\ref{asymetric}). In (\ref{deltagimunu}) the anti-selfdual forms
(the \hika forms) are certainly non-normalizable; being covariantly
constant, they tend to a constant at infinity.
Thus the behaviour of $\delta g_{\mu\nu}=s_{\mu}^{\rho} a_{\rho\nu}$
at infinity is determined by the behaviour of $s_{\mu}^{\rho}$. The
diamonds (\ref{diamonds}) show that the bad-behaved self-dual
two-forms $s_{\mu}^{\rho}$ are three, so we get
\begin{equation}\#\,\,{\rm traceless}\,\,{\bf L}^2\,\,{\rm
deformations}=3\,\tau\, .\label{metricdef2}\end{equation}
We will see that this number has a particularly clear origin in the
construction of the ALE spaces of section \ref{construction}. However
one still has to disregard those deformations that can be readsorbed by means
of diffeomorphisms. One must not consider zeromodes of the
form $\delta g_{\mu\nu}=\nabla_{(\mu}\xi_{\nu)}$ for some vector field
$\xi^{\mu}$. As shown by Hawking and Pope in \cite{hawkingpope}, any
such vector field $\xi^{\mu}$ should tend to one of the $SO(4)$ Killing
vectors of the $S^3$-boundary that commute with the action of $\Gamma$,
given in eq.(\ref{matrixaction},\ref{grouptheory16}).
The generators of $SU(2)_R$ survive for all the possible groups
$\Gamma$; in the case $A_{k-1}, k>2$
(Multi-Eguchi-Hanson metric) also the diagonal generator of $SU(2)_L$
commutes with $A_{k-1}$, and in the Eguchi-Hanson case ($A_1$) all the six
generators of $SO(4)$ commute with $C_2$.

Of course, we must exclude the true Killing vectors of the ALE metric
(that, by definition, do not give rise to any deformation). The ALE
metrics admit one Killing vector in all cases except the Eguchi-Hanson
instanton, which have four of them, as can be seen also from the
explicit form of the metric, eq.(\ref{ehmetric1}).
Summarizing, we have:
\vskip 0.5cm
\begin{center}
\begin{tabular}{c|c|c|c|c|c|c}
      & $C_2$ (E-H) & $A_{k-1}$ & $D_{k+2}$ & $E_6$ & $E_7$ & $E_8$ \\ \hline
\# of defs. & 1 & $3k-6$ & $3k+4$ & 16 & 19 & 22 \\
\end{tabular}
\end{center}

\section{Construction of ALE spaces}\label{construction}
\underline{\sl \hika quotients}
\vskip 0.2cm
Consider a compact Lie group $G$ acting on a \hika manifold $\cS$ of
real dimension $4n$ by means of Killing vector fields $\bx$ holomorphic
with respect to the three complex structures of $\cS$; then these vector
fields preserve also the \hika forms
\footnote{${\cal L}_{\scriptscriptstyle\bx}$ and
$i_{\scriptscriptstyle\bx}$ denote respectively the Lie derivative along
the vector field $\bx$ and the contraction (of forms) with it.}:
\begin{equation}
\left.\begin{array}{l}
{\cal L}_{\scriptscriptstyle\bx}g = 0 \leftrightarrow
\nabla_{(\mu}X_{\nu)}=0 \\
{\cal L}_{\scriptscriptstyle\bx}\cJ i = 0 \,\, , \,i =1,2,3\\
\end{array}\right\} \,\,\Rightarrow\,\, 0={\cal L}_{\scriptscriptstyle\bx}
\Omega^i = i_{\scriptscriptstyle\bx} d\Omega^i+d(i_{\scriptscriptstyle\bx}
\Omega^i) = d(i_{\scriptscriptstyle\bx}\Omega^i)\, .
\label{holkillingvectors}
\end{equation}
If $\cS$ is simply connected, $d(i_{\bx}\Omega^i)=0$ implies the existence
of three functions $\mu_i^{\bx}$ such that $d\mu_i^{\bx}=
i_{\scriptscriptstyle\bx}\Omega^i$. The functions $\mu_i^{\bx}$ are
defined up to a constant,
which can be arranged so to make them equivariant: $\bx \mu_i^{\bf Y} =
\mu_i^{[\bx,{\bf Y}]}$.

The $\{\mu_i^{\bx}\}$ constitute then a {\it momentum
map}. This can be regarded as a map $\mu: \cS \rightarrow \r^3\otimes
\cG^*$, where $\cG^*$ denotes the dual of the Lie algebra $\cG$ of the
group $G$. Indeed let $x\in \cG$ be the element corresponding to the Killing
vector $\bx$; then for a given $m\in \cS$, $\mu_i(m) :x\longmapsto
\mu_i^{\bx}(m)\in\c$ is a linear functional on $\cG$. In practice,
expanding $\bx =X_A {\bf k}^A$ in a basis of Killing vectors ${\bf k}^{A}$
such that $[{\bf k}^A,{\bf k}^B]=C^{AB}_{\hskip 12pt C} {\bf k}^C$, where
$C^{AB}_{\hskip 12pt C} {\bf k}^C$ are the structure constants of $\cG$,
we have also $\mu_i^{\bx}=X_A \mu_i^A$, $i=1,2,3$;
the $\{\mu_i^A\}$ are the components of the momentum map.

The {\it \hika quotient} \cite{hklr} provides a way to construct from
$\cS$ a
lower-dimensional \hika manifold $\cM$, as follows.
Let $\cZ\subset \cG^*$ be
the dual of the centre of $\cG$. For each $\zeta\in \r^3\otimes \cZ$ the
level set of the momentum map
\begin{equation}
\cN \equiv\bigcap_{i} \mu_i^{-1}(\zeta^i) \subset \cS ,
\label{levelsetdef}
\end{equation}
which has dimension $\,\dim\,\cN=\dim\,\cS -3\,\,\dim\,G$,
is left invariant by the action of $G$, due to the equivariance of
$\mu$. It is thus possible to take the quotient
\[\cM =\cN/G.\]
$\cM$ is a manifold of dimension ${\rm dim}\cM={\rm dim}\cS
-4\, {\rm dim}G$ as long as the action of $G$ on $\cN$ has no fixed points.
The three two-forms $\rho^i$ on $\cM$, defined naturally via
the restriction to $\cN\subset\cS$ of the $\Omega^i$ and the quotient
projection from $\cN$ to $\cM$, turn out to be actually \hika
forms on $\cM$.

For future use, it is important to note that, once chosen $\cJ 3$ as
the preferred complex structure, the momentum maps $\mu_{\pm}=\mu_1\pm i\mu_2$
are holomorphic (resp. antiholomorphic).

%

The standard use of the \hika quotient is that of obtaining non trivial
\hika manifolds starting from a
flat $4n$ real-dimensional manifold $\r^{4n}$ acted on by a suitable
group G generating triholomorphic isometries \cite{hklr,lr}. This is the
way it is used also in the construction of ALE manifolds.

The manifold $\r^{4n}$ can be given a quaternionic structure, and the
corresponding quaternionic notation is sometimes convenient. For $n=1$
one has the flat quaternionic space \label{quater}
$\quat\stackrel{\rm def}{=}\left(\r^4,\left\{J^i\right\}\right)$ . We represent
its elements
\[ q\in\quat=x+i y+j z+k t=x^0+x^i J^i,\hskip 15pt x,y,z,t\in\r\]
realizing the quaternionic structures $J^i$ by means of Pauli matrices:
 $J^i=i\left(\sigma^i\right)^T$. Thus
\begin{equation}
q=\twomat{u}{i\bar v}{iv}{\bar u} \hskip 1cm\longrightarrow\hskip 1cm
\bar q=\twomat{\bar u}{-i\bar v}{-iv}{u}
\label{singlequaternion}
\end{equation}
where $u=x^0+ix^3$ and $v=x^1+ix^2$.
The euclidean metric on $\r^4$ is retrieved as $d\bar q\otimes dq=ds^2
\un$. The \hika forms are grouped into a quaternionic two-form
\begin{equation}
\Theta=d\bar q\wedge dq\,\,\stackrel{def}{=}\,\,\Omega^i J^i=
\twomat{i\Omega^3}{i\Omega^+}{i\Omega^-}{-i\Omega^3}\,\, .
\label{thetaform}
\end{equation}
For generic $n$, we have the space $\quat^n$, of elements
\begin{equation}
q=\twomat{u^a}{i\bar v^a}{i v_a}{\bar u_a} \hskip 1cm
\longrightarrow\hskip 1cm \bar q=\twomat{\bar u_a}{-i\bar v^a}
{-iv_a}{u^a}\hskip 1cm
\begin{array}{l}u^a,v_a\in\c^n\\a=1,\ldots n\end{array}
\label{multiquaternion1}
\end{equation}
Thus $d\bar q\otimes dq=ds^2 \un$ gives $ds^2=d\bar
u_a\otimes du^a+d\bar v^a\otimes dv_a$ and the \hika forms are grouped
into the obvious generalization of the
quaternionic two-form eq.(\ref{thetaform}):
$\Theta=\sum_{a=1}^n d\bar
q_a\wedge dq^a=\Omega^i J^i$, leading to $\Omega^3=2i\partial\bar\partial K$
where the \ka potential $K$ is $K=\um\left(\bar u_a u^a + \bar v^a
v_a\right)$,
and to $\Omega^+=2 i du^a\wedge dv_a$,
$\Omega^-=\left(\Omega^+\right)^*$.

Let $\left(T_A\right)^a_b$ be the antihermitean generators of a compact
Lie group G in its  $n\times n$ representation. A triholomorphic action
of $G$ on $\quat^n$ is realized by the Killing vectors of components
\begin{equation}
X_A=\left(\hat T_A\right)^a_b q^b {\partial\over\partial q^a}+\bar q_b
\left(\hat T_A\right)^b_a {\partial\over\partial \bar q_a}\hskip 1cm ;
\hskip 1cm \left(\hat T_A\right)^a_b=\twomat{\left(T_A\right)^a_b}{{\bf
0}}{{\bf 0}}{\left(T^*_A\right)^a_b}\,\, .
\label{triholoaction}
\end{equation}
Indeed one has $\cL_{\scriptscriptstyle \bx}\Theta=0$.
The corresponding components of the momentum map are:
\begin{equation}
\mu^A=\bar q^a\twomat{\left(T_A\right)^a_b}{{\bf
0}}{{\bf 0}}{\left(T^*_A\right)^a_b}q^b +\twomat{c}{\bar b}{b}{-i c}
\label{momentumcomponents}
\end{equation}
where $c\in\r,\,b\in\c$ are constants.
\vskip 0.2cm\noindent
\underline {\sl Kronheimer construction}
\vskip 0.2cm
The \hika quotient is performed on a suitable flat \hika space $\cS$
that now we define. Given any finite subgroup of $SU(2)$, $\Gamma$, consider a
space $\cP$ whose elements are two-vectors of $|\Gamma |\times |\Gamma |$
complex matrices: $p\in \cP =\left(A, B\right)$.
The action of an element $\gamma\in \Gamma$ on the points of $\cP$ is
the following:
\begin{equation}
\twovec{A}{B} \hskip 10pt \stackrel{\gamma}{\longrightarrow}\hskip 10pt
\twomat{u_{\gamma}}{i\bar v_{\gamma}}{iv_{\gamma}}{\bar u_{\gamma}}
\twovec{R(\gamma)AR(\gamma^{-1})}{R(\gamma)BR(\gamma^{-1})}
\label{gammaactiononp}
\end{equation}
where the twodimensional matrix in the r.h.s. is the realization of
$\gamma$ in the defining representation $\cQ$ of $\Gamma$, while $R(\gamma)$
is the regular, $|\Gamma|$-dimensional
representation, defined in section \ref{grouptheory}.
This transformation property identifies $\cP$, from the
point of view of the representations of $\Gamma$, as $\cQ\otimes {\rm End}(R)$.
The space $\cP$ can be given a quaternionic structure,
representing its elements as ``quaternions of matrices'':
\begin{equation}
p\in\cP=\twomat{A}{iB^{\dagger}}{iB}{A^{\dagger}}\hskip 1cm A,B\in {\rm
End}(R)\,\, .
\label{pquaternions}
\end{equation}
The space $\cS$ is the subspace of $\Gamma$-invariant elements in $\cP$:
\begin{equation}
\cS\eqdef\left\{p\in\cP / \forall \gamma\in\Gamma, \gamma\cdot p =
p\right\}\,\, .
\label{mdefinition}
\end{equation}
Explicitly the invariance condition  reads:
\begin{equation}
\twomat{u_{\gamma}}{i\bar v_{\gamma}}{iv_{\gamma}}{\bar u_{\gamma}}
\twovec{A}{B} \hskip 5pt =\hskip 5pt
\twovec{R(\gamma^{-1})AR(\gamma)}{R(\gamma^{-1})BR(\gamma)}\,\, .
\label{invariancecond}
\end{equation}
The space $\cS$ is elegantly described for all $\Gamma$'s using the
associated Dynkin diagram.

A two-vector of matrices can be thought of also as a matrix of
two-vectors: that is, $\cP=\cQ\otimes{\rm Hom}(R,R)={\rm
Hom}(R,\cQ\otimes R)$. Decomposing into irreducible
representations the regular
representation, $R=\bigoplus_{\nu=0}^{r} n_{\mu} D_{\mu}$,
using eq.(\ref{grouptheory16}) and the Schur's lemma, one gets
\begin{equation}
\cS=\bigoplus_{\mu,\nu} A_{\mu,\nu}{\rm
Hom}(\c^{n_{\mu}},\c^{n_{\nu}})\,\, .
\label{defmuastratta}
\end{equation}
The dimensions of the irrepses,  $n_{\mu}$ are expressed in
Fig.s (\ref{dynfigure1},\ref{dynfigure2}).
{}From eq.(\ref{defmuastratta}) the real dimension of
$\cS$ follows immediately: $\dim\, \cS=\sum_{\mu,\nu}2 A_{\mu\nu}
n_{\mu}n_{\nu}$ implies, recalling that $A=2\un-\bar c$ [see
eq.(\ref{grouptheory16})] and that for the  extended Cartan matrix
$\bar c n =0$, that
\begin{equation}
\dim\, \cS=4\sum_{\mu}n_{\mu}^2= 4 |\Gamma |\,\, .
\label{dimm}
\end{equation}

The quaternionic structure of $\cS$ can be seen by simply writing its
elements as in eq.(\ref{pquaternions}) with $A,B$ satisfying the invariance
condition eq.(\ref{invariancecond}). Then the \hika forms and the
metric are described by $\Theta=\tr (d\bar m\wedge m)$ and
$ds^2\un=\tr(d\bar m\otimes d m)$. The
trace is taken over the matrices belonging to ${\rm End}(R)$ in each
entry of the quaternion.
\vskip 0.2cm
{\sl \underline{Example}}\hskip 4pt
The space $\cS$ can be easily described when $\Gamma$ is
the cyclic group $A_{k-1}$. The order of $A_{k-1}$ is $k$;
the abstract multiplication
table is that of ${\bf Z}_k$. We can immediately read off from it the
matrices of the regular representation; of course, it is sufficient to
consider the representative of the first element $e_1$, as
$R(e_j)=(R(e_1))^j$. One has
\begin{equation}
R(e_1)=\left(\matrix{0&0&\cdots &0&1\cr 1&0&\cdots &0&0\cr 0&1&\cdots
&0&0\cr \vdots &\vdots &\ddots &\vdots &\vdots\cr 0&0&\cdots &1&0}\right)
\label{r1offdiag}
\end{equation}
Actually, the invariance condition eq.(\ref{invariancecond})
is best solved by changing
basis so as to diagonalize the regular representation, realizing
explicitly its decomposition in terms of the $k$ unidimensional
irrepses. Let $\nu=e^{2\pi i\over k}$, so that $\nu^k=1$. The
wanted change of basis is performed by the matrix $S_{ij}={\nu^{-ij} \over
\sqrt{k}}$ , such that $S_{ij}^{-1}={\nu^{ij}\over
\sqrt{k}}=S_{ij}^{\dagger}$. In the new basis $R(e_1)={\rm
diag}(1,\nu,\nu^2,\ldots,\nu^{k-1})$, and so
\begin{equation}
R(e_j)={\rm diag}(1,\nu^j,\nu^{2j},\ldots,\nu^{(k-1)j})\,\, .
\label{rej}
\end{equation}
Eq.(\ref{rej}) displays on the diagonal the representatives of
$e_j$ in the unidimensional irrepses.

The explicit solution of eq.(\ref{invariancecond}) is
given in the above basis by
\begin{equation}
A=\left(\matrix{0&u_0&0&\cdots &0\cr 0&0&u_1&\cdots &0\cr \vdots&\vdots&
\vdots&\ddots &\vdots\cr
\vdots &\vdots &\vdots &\null & u_{k-2}\cr
u_{k-1}&0&0&\cdots &0}\right)
\hskip 0.3cm ;\hskip 0.3cm
B=\left(\matrix{0&0&\cdots &\cdots&v_0\cr v_1&0&\cdots &\cdots &0\cr
0&v_2&\cdots &\cdots &0\cr
\vdots &\vdots &\ddots &\null &\vdots \cr
0&0&\cdots &v_{k-1}&0}\right)
\label{invariantck}
\end{equation}
We see that these matrices are parametrized in terms of $2k$ complex,
i.e. $4k=|A_{k-1}|$ real parameters.

In the $D_{k+2}$ case, where the regular representation is $4k$-dimensional,
choosing appropriately a basis, one can solve analogously
eq.(\ref{invariancecond});
the explicit expressions are somehow space-consuming, so we don't write
them. The essential point is that the matrices $A$ and $B$ no longer
correspond to two distinct set of parameters, the group being
non-abelian.
\vskip 0.2cm
Consider the action of $SU(|\Gamma |)$ on $\cP$ given, using the quaternionic
notation for the elements of $\cP$, by
\begin{equation}
\forall g\in SU(|\Gamma |),
g: \twomat{A}{iB^{\dagger}}{iB}{A^{\dagger}}
\longmapsto \twomat{gAg^{-1}}{igB^{\dagger}g^{-1}}{igBg^{-1}}
{gA^{\dagger}g^{-1}}\,\, .
\label{sunaction}
\end{equation}
It is easy to see that this action is a
triholomorphic isometry of $\cP$:\hskip 10pt $ds^2$ and $\Theta$
are invariant.
Let $F$ be the subgroup of $SU(|\Gamma |)$ which {\it commutes with the
action of $\Gamma$ on $\cP$}, with action described in
eq.(\ref{gammaactiononp}).
Then the action
of $F$ descends to $\cS\subset\cP$ to give a
{\it triholomorphic isometry}: the metric and \hika forms on $\cS$ are just
the restriction of those on $\cP$. It is therefore possible to take the
\hika quotient of $\cS$ with respect to $F$.

Let $\{f_A\}$ be a basis of generators for $\cF$, the Lie algebra of $F$. Under
the infinitesimal action of $f=\un+\lambda^A f_A\in F$,
the variation of $m\in \cS$
is $\delta m= \lambda^A\delta_A m$, with
\begin{equation}
\delta_A m = \twomat{[f_A,A]}{i[f_A,B^{\dagger}]}{i[f_A,B]}
{[f_A,A^{\dagger}]}\,\, .
\label{deltaam}
\end{equation}
The components of the momentum map (see (\ref{momentumcomponents}))
are then given by
\begin{equation}
\mu_A=\tr\,(\bar m\,\delta_A m)\,\,\,\eqdef\,\,\,
\tr\,\twomat{f_A\,\mu_3(m)}{f_A\,\mu_-(m)}{f_A\,\mu_+(m)}{f_A\,\mu_3(m)}
\label{momentummatrix}
\end{equation}
so that the real and holomorphic maps $\mu_3:\cS\rightarrow\cF^*$ and
$\mu_+:\cS\rightarrow\c\times\cF^*$ can be represented as matrix-valued maps:
\footnote{It is easy to see that indeed the matrices
$[A,A^{\dagger}]+[B,B^{\dagger}]$ and $[A,B]$ belong to the Lie algebra
of traceless matrices $\cF$; practically we identify $\cF^*$ with $\cF$
by means of the Killing metric.}
\begin{eqnarray}
\mu_3(m)&=&-i\left([A,A^{\dagger}]+[B,B^{\dagger}]\right)\nonumber\\
\mu_+(m)&=&\left([A,B]\right)\,\, .
\label{momentums}
\end{eqnarray}
Let $\cZ$ be the dual of the centre of $\cF$.
In correspondence of a level $\zeta=\{\zeta^3,\zeta^+\}\in{\bf
R}^3\otimes\cZ$ we can form the \hika quotient
$\cM_{\zeta}\eqdef\mu^{-1}/F$. {\it Varying $\zeta$ and $\Gamma$ every
ALE space can be obtained as $\cM_{\zeta}$}.

First of all, it is not difficult to check that $\cM_{\zeta}$ is
four-dimensional. As for the space $\cS$, there is a nice characterization
of the group $F$ in terms of the extended Dynkin diagram associated
with
$\Gamma$:
\begin{equation}
F=\bigotimes_{\mu} U(n_{\mu})\,\, .
\label{formofF}
\end{equation}
One must however set
the determinant of the elements to be one, since $F\subset SU(|\Gamma |)$.
$F$ has a $U(n_{\mu})$ factor for each dot of the diagram, $n_{\mu}$ being
associated to the dot as in Fig.s \ref{dynfigure1},\ref{dynfigure2}.
$F$ acts on the various
``components'' of $\cS$ [which are in correspondence with the edges of
the diagram, see eq.(\ref{defmuastratta})] as dictated by the structure
diagram. From
eq.(\ref{formofF}) is immediate to derive that $\dim\, F=\sum_{\mu}
n_{\mu}^2 -1 = |\Gamma |-1$. It follows that
\begin{equation}
\dim \cM_{\zeta}=\dim\, \cS -4\, \dim F = 4|\Gamma | -4(|\Gamma
|-1)=4\,\, .
\label{dimxzeta}
\end{equation}
\vskip 0.2cm
{\sl \underline{Example}}\hskip 4pt
The structure of $F$ and the momentum map for its action are very simply
worked out in the $A_{k-1}$ case. An element $f$ of $F$ must commute with the
action of $A_{k-1}$ on $\cP$, eq.(\ref{gammaactiononp}),
where the two-dimensional
representation in the l.h.s. is given in eq.(\ref{grouptheory7}).
Then $f$ must have the form
\begin{equation}
f={\rm diag} (e^{i\varphi_0},e^{i\varphi_1},\ldots ,e^{i\varphi_{k-1}})
\hskip 0.2cm ; \hskip 0.2cm \sum \varphi_{i}=0\,\, .
\label{Fforck}
\end{equation}
Thus $\cF$ is just the algebra of diagonal traceless $k$-dimensional
matrices, which is $k-1$-dimensional. Choose a basis of generators for
$\cF$, for instance $f_1=\diag
(1,-1,\ldots)$,$f_2=\diag (1,0,-1,\ldots)$,$\ldots$,$f_{k-1}=\diag
(1,0,\ldots,-1)$. From eq.(\ref{momentums}) one gets directly the components
of the momentum map:
\begin{eqnarray}
\mu_{3,A}&=&|u^0|^2-|v_0|^2-|u^{k-1}|^2-|v_{k-1}|^2
-|u^A|^2-|v_A|^2+|u^{A-1}|^2-|v_{A-1}|^2\nonumber \\
\mu_{+,A}&=&u^0 v_o - u^{k-1}v_{k-1}- u^A v_A + u^{A-1} v_{A-1}\,\, .
\label{momentummapck}
\end{eqnarray}
\vskip 0.2cm
In order for $\cM_{\zeta}$ to be a manifold, it is necessary that $F$ act
freely on $\mu^{-1}(\zeta)$. This happens or not depending on the value
of $\zeta$.
Again, a simple characterization of $\cZ$ can be given in terms of the
simple Lie algebra $\cG$ associated with $\Gamma$ \cite{kro1}. There exists an
isomorphism between $\cZ$ and the Cartan subalgebra $\cH$ of $\cG$. Thus
we have
\begin{eqnarray}
\dim\, \cZ=\dim\,\cH &=&{\rm rank}\,\cG\nonumber\\
&=&\#{\rm
of\,\,\,non\,\,\,trivial\,\,\,conj.\,\,\,classes\,\,\,in}\,\,\,\Gamma
\,\, .
\end{eqnarray}
The space $\cM_{\zeta}$ turns out to be singular when, under the above
identification $\cZ\sim\cH$, any of the level components $\zeta^i\in {\bf
R}^3\otimes \cZ$ lies on the walls of a Weyl chamber.
In particular, as the point $\zeta^i=0$ for all $i$ is identified with
the origin in the root space, which lies of course on all the walls of
the Weil chambers, {\it the space $\cM_0$ is singular}. Without too much
surprise we will see in a momentum that $\cM_0$ corresponds to the {\it
orbifold limit} $\csg$ of a family of ALE manifolds with boundary at
infinity $S^3/\Gamma$.

To see that this is general, choose the natural basis for the
regular representation $R$, in which the basis
vectors $e_{\delta}$ transform as in eq.(\ref{grouptheory15}).
Define then the space $L\subset \cS$ as follows:
\begin{equation}
L=\left\{\twovec{C}{D}\in\cS\,/\,C,D \,\,{\rm
are\,\,diagonal\,\,in\,\,the\,\,basis\,\,}\left\{e_{\delta}\right\}
\right\}\,\, .
\label{thespacel}
\end{equation}
For every element $\gamma\in\Gamma$ there is a pair of numbers
$(c_{\gamma},d_{\gamma})$
given by the corresponding entries of $C,D$:
$C\cdot e_{\gamma}=c_{\gamma}e_{\gamma}$, $D\cdot e_{\gamma}=d_{\gamma}
e_{\gamma}$. Applying the invariance condition eq.(\ref{invariancecond}),
which is valid since $L\subset\cS$, it results that
\begin{equation}
\twovec{c_{\gamma\cdot\delta}}{d_{\gamma\cdot\delta}}=
\twomat{u_{\gamma}}{i\bar v_{\gamma}}{iv_{\gamma}}{\bar u_{\gamma}}
\twovec{c_{\delta}}{d_{\delta}}\,\, .
\label{orbitofthepair}
\end{equation}
We can identify $L$ with $\c^2$ associating for instance $(C,D)\in L
\longmapsto (c_0,d_0)\in \c^2$. Indeed all the other pairs $(c_{\gamma},
d_{\gamma})$ are determined in terms of eq.(\ref{orbitofthepair}) once
$(c_0,d_0)$ are given. By eq.(\ref{orbitofthepair}) the action
of $\Gamma$ on $L$ induces exactly the action of $\Gamma$ on $\c^2$ that
we considered in (\ref{matrixaction},\ref{grouptheory3}).

It is quite easy to show the following fundamental fact: {\it
each orbit of $F$ in $\mu^{-1}(0)$ meets $L$ in one orbit of $\Gamma$}.
Because of the above identification between $L$ and $\c^2$, this leads
to prove that {\it $X_0$ is isometric to $\csg$}.

In the spirit of the paper, instead of reviewing the proof of these
statements (see \cite{kro1}), we show explicitly the above facts in the
case of the cyclic groups, giving a description which sheds
some light on the {\it deformed} situation; that is we show in which
way a non-zero level $\zeta^+$ for the holomorphic momentum map puts
$\mu^{-1}(\zeta)$ in correspondence with an hypersurface in $\c^3$
defined by a ``potential'' which is a deformation of the one describing
the $\csg$ situation, obtained for $\zeta^+=0$.
\vskip 0.2cm\noindent
\vskip 0.2cm\noindent
\underline{\sl The case $\Gamma$=\ak}
\vskip 0.2cm
We can directly realize $\csg$ as an affine
algebraic surface in $\c^3$ (see eq. (\ref{grouptheory20}))
by expressing the coordinates $x$, $y$ and $z$ of $\c^3$
in terms of the matrices $(C,D)\in L$.
The explicit parametrization of the matrices in ${\cal S}$ in the
$A_{k-1}$case
(which was given in eq.(\ref{invariantck}) in the basis in which the regular
representation $R$ is diagonal), can be
conveniently rewritten in the ``natural'' basis $\left\{e_{\gamma}
\right\}$ via the matrix $S^{-1}$ [see before eq.(\ref{rej})]. The subset
$L$ of diagonal matrices $(C,D)$ is given by
\begin{equation}
C=c_0\, {\rm diag}(1,\nu,\nu^2,\ldots,\nu^{k-1}),\hskip 12pt
D=d_0\, {\rm diag}(1,\nu^{k-1},\nu^{k-2},\ldots,\nu),
\label{unouno}
\end{equation}
where $\nu={\rm e}^{2\pi i\over k}$. This is nothing but the fact that
$\c^2\sim L$. The set of pairs
$\left(\matrix{\nu^m c_0\cr \nu^{k-m}d_0}
\right)$, $m=0,1,\ldots,k-1$ is an orbit of $\Gamma$ in $\c^2$
and determines the corresponding orbit of $\Gamma$ in $L$.
To describe $\c^2 / A_{k-1}$
one needs to identify a suitable set of invariants $(x,y,z)\in \c^3$
such that $xy=z^k$,
namely eq. (\ref{grouptheory20}).
Our guess is
\begin{equation}
x=\det\, C \hskip 12pt ;\hskip 12pt y=\det\, D,
\hskip 12pt ;\hskip 12pt z={1\over k} \tr \,CD.
\label{identifyAk}
\end{equation}
This guess will be confirmed in a moment by the study of the deformed surface.

We know that there is a one-to-one correspondence between the orbits of
$F$ in $\mu^{-1}(0)$ and those of $\Gamma$ in $L$. Let us realize
it explicitly. Choose the basis where $R$ is diagonal. Then $(A,B)
\in {\cal S}$ has the form of eq. (\ref{invariantck}). Now,
the relation $xy=z^k$ (eq. (\ref{grouptheory20}))
holds also true when, in eq. (\ref{identifyAk}), the pair $(C,D)\in L$
is replaced
by an element $(A,B)\in \mu^{-1}(0)$.
To see this, let us describe the elements $(A,B)\in\mu^{-1}(0)$. We have to
equate the right hand sides
of eq. (\ref{momentums}) to zero. We note that
$[A,B]=0$ gives $v_i={u_0v_0\over v_i}$ $\forall i$.
Secondly, $[A,A^\dagger]+[B,B^\dagger]=0$ implies $|u_i|=|u_j|$
and $|v_i|=|v_j|$ $\forall i,j$, i.e.\ $u_j=|u_0|{\rm e}^{i\phi_j}$
and $v_j=|v_0|{\rm e}^{i\psi_j}$. Finally, $[A,B]=0$ implies $\psi_j=\Phi-
\phi_j$ $\forall j$ for a certain phase $\Phi$.
In this way, we have characterized $\mu^{-1}(0)$
and we immediately check that the pair $(A,B)\in\mu^{-1}(0)$
satisfies $xy=z^k$ if $x=\det\,A$, $y=\det\,B$ and $z=(1/k) \, \tr \,AB$.
We are left  with $k+3$ parameters (the $k$ phases
$\phi_j$, $j=0,1,\ldots k-1$,
plus the absolute values $|u_0|$ and $|v_0|$ and the phase $\Phi$).
Indeed ${\rm dim}\,\mu^{-1}(0)={\rm dim}\,{\cal M}-3 \,{\rm dim}\,F=
4|\Gamma|-3(|\Gamma|-1)=|\Gamma|+3$, where $|\Gamma|={\rm dim}\,\Gamma=k$.

Now we perform the quotient of $\mu^{-1}(0)$
with respect to $ F $.
Given a set of phases $f_i$ such that $\sum_{i=0}^{k-1}f_i=0\,
{\rm mod} \, 2\pi$ and given $f={\rm diag}
({\rm e}^{if_0},{\rm e}^{if_1},\ldots,{\rm e}^{if_{k-1}})\in  F $,
the orbit of $ F $ in $\mu^{-1}(0)$ passing through
$\left(\matrix{A\cr B}\right)$ is given by
$\left(\matrix{fAf^{-1}\cr fBf^{-1}}\right)$. Choosing
$f_j=f_0+j\psi+\sum_{n=0}^{j-1}\phi_n$, $j=1,\ldots,k-1$, with
$\psi=-{1\over k}\sum_{n=0}^{k-1}\phi_n$, and $f_0$ determined by
the condition $\sum_{i=0}^{k-1}f_i=0\, {\rm mod} \, 2\pi$, one has
\begin{equation}
fAf^{-1}=a_0\left(\matrix{
0 & 1 & 0 & \ldots & 0\cr
0 & 0 & 1 & \ldots & 0\cr
  & \ldots  & & \dots  &  \cr
0 & 0  &  \dots & 0 & 1 \cr
1 & 0 & 0 & \dots & 0 \cr
}\right)\, ,  \hskip 10pt
fBf^{-1}=b_0\left(\matrix{
0 & 0  &  \dots & 0 & 1 \cr
1 & 0 & 0 & \dots & 0 \cr
0 & 1 & 0 & \ldots & 0\cr
  & \ldots  & & \dots  &  \cr
0 & \ldots & 0 & 1 & 0\cr
}\right)
\label{op}
\end{equation}
where $a_0=|u_0|{\rm e}^{i\psi}$ and $b_0=|v_0|{\rm e}^{i(\Phi-\psi)}$.
Since the phases $\phi_j$ are determined modulo $2\pi$, it follows that
$\psi$ is determined modulo $2\pi\over k$. Thus we can say
$(a_0,b_0)\in {\csg}$.
This is the one-to-one correspondence between
$\mu^{-1}(0)/F$ and $\c^2/\Gamma$.

We now derive the deformed relation between the invariants $x,y,z$.
It fixes the correspondence between the resolution of singularity
performed in the momentum map approach and the resolution performed on the
hypersurface $xy=z^k$ in $C^3$. To this purpose, we focus on the holomorphic
part of the momentum map, i.e.\ on the equation $[A,B]=\Lambda_0$, where
$\Lambda_0={\rm diag}(\lambda_0,\lambda_1,\lambda_2,\ldots,\lambda_{k-1})\in
{\cal Z}\otimes\c^2$ with $\lambda_{0}=-\sum_{i=1}^{k-1}\lambda_i$.
Recall expression (\ref{invariantck}) for the matrices $A$ and $B$.
Calling $a_i=u_iv_i$, $[A,B]=\Lambda_0$ implies that
$a_i=a_0+\lambda_i$ for $i=1,\ldots,k-1$. Now,
let $\Lambda={\rm diag}(\lambda_1,\lambda_2,\ldots,\lambda_{k-1})$.
We have
\begin{equation}
xy=\det A \, \det B=
a_0\, \Pi_{i=1}^{k-1}(a_0+\lambda_i)=a_0^k\,\det \left(1+{1\over a}
\Lambda\right)=\sum_{i=0}^{k-1}a_0^{k-i}S_i(\Lambda).
\label{def1}
\end{equation}
The $S_i(\Lambda)$ are the symmetric polynomials in the eigenvalues
of $\Lambda$. They are defined by the relation $\det (1+\Lambda)=
\sum_{i=0}^{k-1}S_i(\Lambda)$ and are given by
$S_i(\Lambda)=\sum_{j_1<j_2<\cdots<j_i}\lambda_{j_1}\lambda_{j_2}
\cdots \lambda_{j_i}$. In particular, $S_0=1$ and $S_1=\sum_{i=1}^{k-1}
\lambda_i$.
Define $S_k(\Lambda)=0$, so that $xy=
\sum_{i=0}^{k}a_0^{k-i}S_i(\Lambda)$, and note that
\begin{equation}
z={1\over k}\tr AB=a_0+{1\over k}S_1(\Lambda).
\end{equation}
Then the desired deformed relation between $x$, $y$ and $z$
is obtained by substituting
$a_0=z-{1\over k}S_1$ in (\ref{def1}), thus obtaining
\begin{eqnarray}
&&xy=\sum_{m=0}^k\sum_{n=0}^{k-m}\left(\matrix{k-m\cr n}\right)
\left(-{1\over k}S_1\right)^{k-m-n}S_m z^n=\sum_{n=0}^k
t_n z^n.\\
&&\Longrightarrow\hskip 10pt t_n=\sum_{m=0}^{k-n}\left(\matrix{k-n\cr m}\right)
\left(-{1\over k}S_1\right)^{k-m-n} S_n .
\label{def2}
\end{eqnarray}
Notice in particular that $t_k=1$ and $t_{k-1}=0$, i.e.\
$xy=z^k+\sum_{n=0}^{k-2}t_nz^n$, which
means that the deformation proportional to $z^{k-1}$ is absent.
This establishes a clear correspondence between the momentum
map construction and the polynomial ring ${C[x,y,z]\over \partial W}$
where $W(x,y,z)=xy-z^k$ (compare with eq. ({\ref{grouptheory21})).
Moreover, note that we have only used one
of the momentum map equations, namely $[A,B]=\Lambda_0$.
The equation $[A,A^\dagger]+[B,B^\dagger]=\Sigma$ has been completely
ignored. This means that the deformation of the complex structure
is described by the parameters $\Lambda$, while the parameters
$\Sigma$ describe the deformation of the complex structure.

The relation (\ref{def2}) can also be written in a simple factorized form,
namely
\begin{equation}
xy=\Pi_{i=0}^{k-1}(z-\mu_i),
\end{equation}
where
\begin{eqnarray}
\mu_i&=&{1\over k}(\lambda_1+\lambda_2+\cdots +\lambda_{i-1}
-2\lambda_i+\lambda_{i+1}+\cdots+\lambda_k), \,\,\,\,
i=1,\ldots,k-1\nonumber\\
\mu_0&=&-\sum_{i=1}^k\mu_i={1\over k}S_1.
\end{eqnarray}
\vskip 0.2cm\noindent
\underline{The case $\Gamma$=\dk}
\vskip 0.2cm
The case $\Gamma$=\dk\ cannot be treated with the algebraic simplicity of
the previous one. Nevertheless, we can give an ansatz for the expressions
of $x$, $y$ and $z$ in terms of the matrices $(A,B)\in \mu^{-1}(\zeta)/F$.
This ansatz surely works for the undeformed case $\zeta=0$, because it can
be checked via the correspondence between $\mu^{-1}(0)/F$ and $\c^2/\Gamma$
that permits to manage with diagonal matrices $C$ and $D$ instead of $A$
and $B$. Let (compare with (\ref{grouptheory29}))
\begin{eqnarray}
x&=&{i\over 8k}\tr\,[A^{2k+1}B-(-1)^k A\, B^{2k+1}],\nonumber\\
y&=&{i\over 8k}\tr\,[A^{2k}+(-1)^k B^{2k}]\nonumber\\
z&=&-{1\over 16k}\tr \,\{A,B\}^2.
\end{eqnarray}
In the undeformed case
$\zeta=0$, the relation $[A,B]=0$ shows that one cannot fix $z$
unambiguously, because expressions proportional
to $\tr\, A^2B^2$ or $\tr\, (AB)^2$ are equally allowed. To resolve the
ambiguity, we have worked out the deformation in the simplest case, namely
$k=1$. $A$ and $B$ are $4\times 4$ matrices. In a suitable basis
they have the form
\begin{equation}
\matrix{A=\left(\matrix{0&a&0&b\cr c&0&d&0\cr
0&e&0&f\cr g&0&h&0}\right),&B=i\left(\matrix{0&-f&0&e\cr
-h&0&-g&0\cr 0&-b&0&a\cr d&0&c&0}\right)
}.
\end{equation}
In our case the explicit form of the momentum map equation is
$\mu_+(m)\equiv [A,B]=\Lambda$ [see eq.(\ref{momentums})], that we write as
\begin{equation}
[A,B]=i\left(\matrix{l_1&0&l_2&0\cr 0&-l_1&0&l_3\cr
l_2&0&l_1&0\cr 0&-l_3&0&-l_1}\right).
\end{equation}
Then we have
\begin{equation}
x^2+y^2z+z^2+t_1+t_2 y + t_3 y^2+ t_4 z =0,
\end{equation}
where
\begin{eqnarray}
t_1&=&-{1\over 16}\left[l_2^2l_3^2-{1\over 4}(2l_1^2-l_2^2+l_3^2)\right],
\nonumber\\
t_2&=&-{i\over 4}l_1l_2l_3,\nonumber\\
t_3&=&{1\over 8}(l_2^2-l_3^2),\nonumber\\
t_4&=&-{1\over 4}l_1^2.
\end{eqnarray}
Note the presence of both $y^2$ and $z$ in the deformed relation, although
one vanishing relation of the chiral ring says that they are proportional.
One can make the $y^2$-term disappear by simply performing
a $l$-dependent translation of $z$.

\section{The orbifold $\c^2/\Gamma$ conformal field theory}
\label{cft}
Now we address the problem of constructing the $(4,4)$ conformal field theory
associated with an ALE instanton. This can be explicitly done in the
orbifold limit $\cM_0=\csg$, corresponding to $\left\{\zeta^i=0\right\}$.
The $(4,4)$ theories associated with the smooth manifolds $\cM_{\zeta}$
can be obtained from the orbifold theory by perturbing it with the
moduli operators associated with the elements of the ring $\c
[x,y,z]/\partial W$.

After briefly reviewing  the structure of the $N=4$ algebra and its
possible representations at $c=6$, we give a description of the $(4,4)$
orbifold theory in terms of primary fields and OPEs. Next we
consider the structure of the corresponding partition function, and
we analyse it in terms of $N=4$ characters.

The $N=4$ algebra is described in terms of OPEs as follows:
\begin{eqnarray}
\label{n=4ope}
T(z)T(w)&=&{c\over 2}{1\over (z-w)^4}+{2T(w)\over (z-w)^2}+
{\partial T(w)\over z-w}\nonumber\\
T(z)\G^a(w)&=& {3\over 2}{\G^a(w)\over (z-w)^2}+{\partial \G^a\over z-w}
\hskip 2pt ; \hskip 2pt
T(z)\bar\G^a(w)= {3\over 2}{\bar\G^a(w)\over (z-w)^2}+
{\partial \bar\G^a\over z-w}
\nonumber\\
T(z)A^i(w)&=&{A^i(w)\over (z-w)^2}+{\partial A^i(w)\over z-w}\nonumber\\
A^i(z)\G^a(w)&=& {1\over 2}{\G^b(w) (\sigma^i)^{ba}\over z-w}
\hskip 4pt ;\hskip 4pt
A^i(z)\bar\G^a(w)=- {1\over 2}{\bar\G^b(w) (\sigma^i)^{ab}\over z-w}
\nonumber\\
\G^a(z)\bar\G^b(w)&=&{2\over 3}c {\delta^{ab}\over (z-w)^3}+
{4 (\sigma_i^*)^{ab}A^i(w)\over (z-w)^2}+2{\delta^{ab}T(w)+
\partial A^i(w)(\sigma^*_i)^{ab}\over z-w}\nonumber\\
A^i(z)A^j(w)&=&{1\over 12}c {\delta^{ij}\over (z-w)^2}+i
\varepsilon^{ijk}{A^{(k)}(w)\over z-w}\,\, .
\end{eqnarray}
Note that in the above OPEs and in all the following ones the equality sign
means equality up to regular terms.
In general, the central charge $c$ is an integer multiple of
$6$ in a unitary theory, but we shall only be interested in the case $c=6$.

To discuss the structure of the
representations \cite{eguchi}, using the highest-weight method,
it is convenient to rewrite the $N=4$ algebra (\ref{n=4ope}) in terms of modes:
\begin{eqnarray}
\label{n=4modes}
&&[L_m,L_n]=(m-n)L_{m+n}+{1\over 2}km(m^2-1)\delta_{m+n,0}\nonumber\\
&&[L_m,\cG^a_r]=({1\over 2}m-r)\cG^a_{m+r},\ \ \
[L_m,\bar \cG^a_s]=({1\over 2}m-s)
\bar \cG^a_{m+s}\nonumber\\
&&[L_m,A^i_n]=-nA^i_{m+n}\nonumber\\
&&[A^i_m,\cG^a_r]={1\over 2}\sigma^i_{ba}\cG^b_{m+r},\ \ \  [A^i_m,\bar
\cG^a_s]=
-{1\over 2}\sigma^{i*}_{ba}\bar \cG^b_{m+s}\nonumber\\
&&\{\cG^a_r,\cG^b_s\}=\{\bar \cG^a_r,\bar \cG^b_s\}=0\nonumber\\
&&\{\cG^a_r,\bar \cG^b_s\}=2\delta^{ab}L_{r+s}-2(r-s)
\sigma^i_{ab}A^i_{r+s}+
{1\over 2}k(4r^2-1)\delta_{r+s,0}\delta_{ab}
\nonumber\\
&&[A^i_m,A^j_n]=i\epsilon_{ijk}A^k_{m+n}+{1\over 2}km\delta_{m+n,0}\delta^{ij}
\end{eqnarray}
 where indices $r,s$ take integral values in the Ramond sector (R) and
half-integer values in the Neveu-Schwarz (NS) sector. The value
of the central charge being $c=6k$, only the case $k=1$ is relevant to
our discussion, as already stated. Furthermore we can restrict ourselves
to the NS sector, since the Ramond sector can be reached by
spectral-flow.\par
The highest-weight states of the N=4 algebra are defined by the conditions:
\begin{eqnarray}
&&L_n|h,l\rangle = \cG^a_r|h,l\rangle =\bar \cG^b_s|h,l\rangle =
A^i_n|h,l\rangle
= 0,\ \ \  n\geq 1,\ \  r,s\geq {1\over 2}\nonumber\\
&&A^+_0|h,l\rangle = 0\nonumber\\
&&L_0|h,l\rangle = h|h,l\rangle ,\ \  T^3_0|h,l\rangle
=l|h,l\rangle\,\, .
\end{eqnarray}
Unitarity puts the restriction $h\geq l$. There exist two classes of unitary
representations of the N=4 algebra: the {\it long} representations
\begin{equation}
h>l,\ \ \ \  l=0,{1\over 2},...,{1\over 2}(k-1)
\end{equation}
and the {\it short} ones
\begin{equation}
h=l,\ \ \ \  l=0,{1\over 2},...,{1\over 2}\,\, .
\end{equation}
The short representations exist when $h$ saturates the unitary bound $h=l$ and
the long representations decompose in short ones in the limit in which $h$
reaches $l$. The unitary bound $h\geq l$ is the $N=4$ transcription
of the analogous $N=2$ bound $h\ge |q|/2$.
The short representations are in fact defined to obey the condition
\begin{equation}
\cG^2_{-{1\over 2}}|h,l\rangle = \bar \cG^1_{-{1\over 2}}|h,l\rangle = 0
\end {equation}
which is in fact equivalent to $h=l$ by commutation relations. In other terms,
$|h,h\rangle$ can be constructed using the chiral fields of the corresponding
N=2 algebra.
Starting from the highest-weight state we can try to close
an N=4 superconformal representation
by repeated application of the generators. The result
can be conveniently retrieved and expressed in terms of OPEs, as
follows.

Note that for $c=6$ there are only two type of short representations:
the $|0,0\rangle$ case, corresponding only to the identity in a unitary
conformal theory, and the $|{1\over 2},{1\over 2}\rangle$ case, the only
non trivial one.

Using the notation introduced in section \ref{intro},
the multiplet of a short representation is
\begin{equation}
\left(\Psi^a\left[\matrix{1/2\cr 1/2}\right],
\Phi\left[\matrix{1\cr 0}\right],
\Pi\left[\matrix{1\cr 0}\right]\right).
\end{equation}
It is characterized by the following OPEs with the supercurrents:
\begin{eqnarray}
\G^a(z)\Psi^b(w)&=&\delta^{ab}{\Phi(w)\over z-w}
\hskip 12pt ;\hskip 12pt
\bar\G^a(z)\Psi^b(w)=\epsilon^{ab}{\Pi(w)\over z-w}\nonumber\\
\G^a(z)\Pi(w)&=&-2\varepsilon^{ab}
\partial_w\left({\Psi^a\over z-w}\right)
\hskip 8pt ;\hskip 8pt
\bar\G^a(z)\Phi(w)=2\partial_w\left({\Psi^a\over z-w}\right).\nonumber\\
\label{masslessOPE2}
\end{eqnarray}
The first two OPEs define the short representations,
while the other OPEs are consequences of the first two OPEs.

This can be seen by using two Jacobi-like
identities, that can be written as
\begin{eqnarray}
&&\oint {z^mdz\over 2\pi i}\oint {\zeta^nd\zeta\over 2\pi i}
D^{\pm}(z)\cdot(G(\zeta)\cdot {\cal O}(w))=\nonumber\\
&&=\oint {z^mdz\over 2\pi i}\oint {\zeta^nd\zeta\over 2\pi i}
\Big((D^{\pm}(z)\cdot G(\zeta))\cdot {\cal }O(w)\pm
G(\zeta)\cdot(D^{\pm}(z)\cdot {\cal O}(w))\Big)
\label{jacobi1}
\end{eqnarray}
$\forall m$, $\forall n$ and $\forall {\cal O}(w)$, with
$D^{\pm}(z)$ bosonic (resp. fermionic) and $G(\zeta)$ fermionic.
The dot means OPE expansion. We specify the
order in which the OPE's are to be performed instead of specifying the
equivalent information on the integration contours.
One can use the cases
$m,n=0,1$ to extract information about simple and double poles.
Explicitly, set
\begin{eqnarray}
\varepsilon^{ab}\Phi(w)&=&\oint {d\zeta\over 2\pi i}{\cal G}^a(\zeta)
\Psi^b(w),\nonumber\\
\delta^{ab}\Pi(w)&=&\oint {d\zeta\over 2\pi i}\bar{\cal G}^a(\zeta)
\Psi^b(w).
\end{eqnarray}
With ${\cal O}(w)=\Psi^b(w)$ and $G(\zeta)=
{\cal G}^a(\zeta)$ or $\bar{\cal G}^a(\zeta)$,
the identities (\ref{jacobi1}) can be used alternatively
to check the conformal weight of $\Phi$ and $\Pi$
(by choosing $D^+(z)=T(z)$),
to check the $SU(2)$ representation (with $D^+(z)=A^i(z)$) and finally
to check the other
OPEs of eq. (\ref{masslessOPE2})
($D^-(z)={\cal G}^a(\zeta)$ or $\bar{\cal G}^a(\zeta)$).

The multiplet of a long representations is instead (see Fig.
\ref{repfigure}).
\begin{equation}
\left(\Omega\left[\matrix{h\cr 0}\right],
\Phi^a\left[\matrix{h+1/2\cr 1/2}\right],
\bar \Phi^a\left[\matrix{h+1/2\cr 1/2}\right],
\Gamma\left[\matrix{h+1\cr 0}\right],
\bar\Gamma\left[\matrix{h+1\cr 0}\right],
\Sigma\left[\matrix{h+1\cr 0}\right]\right),
\end{equation}
with $h>0$.
Its OPEs with the supercurrents are
\begin{eqnarray}
\G^a(z)\Omega(w)&=&{\Phi^a(w)\over z-w}\hskip 12pt ; \hskip 12pt
\bar\G^a(z)\Omega(w)\,\,=\,\,{\bar\Phi^a(w)\over z-w}\nonumber\\
\G^a(z)\Phi^b(w)&=&\varepsilon^{ab}{\Gamma(w)\over z-w}
\hskip 12pt ; \hskip 12pt
\bar\G^a(z)\bar\Phi^b(w)=\varepsilon^{ab}{\bar\Gamma(w)\over z-w}\nonumber\\
\bar \G^a(z)\Phi^b&=&2h\delta^{ab}{\Omega(w)\over (z-w)^2}+
\delta^{ab}{\partial\Omega(w)\over z-w}+{1\over 2}\delta^{ab}
{\Sigma(w)\over z-w}-8h{(\sigma_i)^{ab}A^i(w)\Omega(w)\over z-w}
\nonumber\\
\G^a(z)\bar\Phi^b&=&2h\delta^{ab}{\Omega(w)\over (z-w)^2}+
\delta^{ab}{\partial\Omega(w)\over z-w}-{1\over 2}\delta^{ab}
{\Sigma(w)\over z-w}+8h{(\sigma^*_i)^{ab}A^i(w)\Omega(w)\over z-w}
\nonumber\\
\G^a(z)\bar \Gamma(w)&=&2(h+1)\varepsilon^{ab}{\bar\Phi^b\over (z-w)^2}
+2 \varepsilon^{ab}{\partial\bar\Phi^b\over z-w}+2h\varepsilon^{ab}
{\Delta\bar\Phi^b\over z-w}\nonumber\\
\bar\G^a(z)\Gamma(w)&=&2(h+1)\varepsilon^{ab}{\Phi^b\over (z-w)^2}
+2 \varepsilon^{ab}{\partial\Phi^b\over z-w}+2h\varepsilon^{ab}
{\Delta\Phi^b\over z-w}\nonumber\\
\G^a(z)\Sigma(w)&=&2(h+1){\Phi^a\over (z-w)^2}
+2 {\partial\Phi^a\over z-w}+2h{\Delta\Phi^a\over z-w}\nonumber\\
\bar\G^a(z)\Sigma(w)&=&-2(h+1){\bar\Phi^a\over (z-w)^2}
-2 {\partial\bar\Phi^a\over z-w}-2h{\Delta\bar\Phi^a\over z-w}
\label{massivegen}
\end{eqnarray}
where $\Delta\Phi^a(w)=\lim_{w^\prime\rightarrow w}
[4 (\sigma_i)^{ab}A^i(w)\Phi^b(w^\prime)-2\G^a(w)\Omega(w^\prime)]$.
As in the case of the short representations, the first two OPEs
are assumptions. They define the highest weight operator $\Omega$
of the long representation.
All the other OPEs are consequences of the first two and the Jacobi identities
(\ref{jacobi1}) as in the massless case.

The structure of the above OPEs is well summarized in Figure
\ref{repfigure} that can be seen also as describing the algebra closed
on the representations by the zero-modes of the generators.

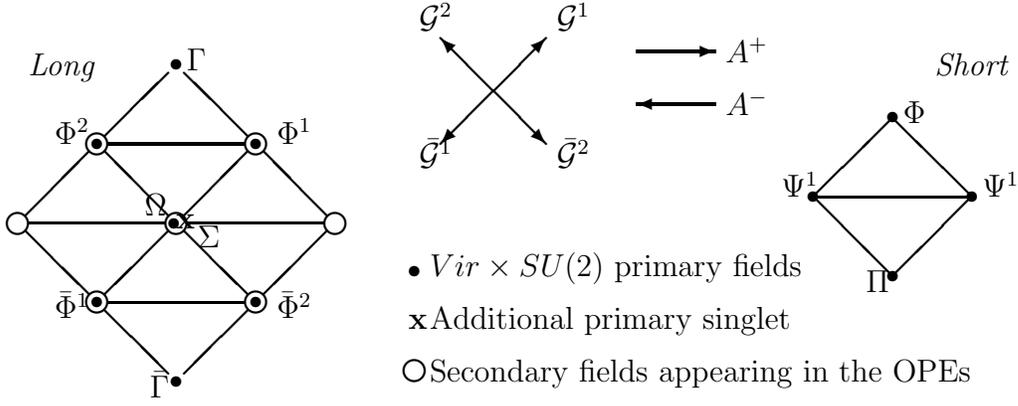
\begin{figure}
\label{repfigure}
\begin{picture}(290,130)(-70,-70)
\thicklines
\put(-56,56){\it Long}
\multiput(-60,0)(60,0){3}{\circle{8}}
\multiput(-30,30)(60,0){2}{\circle*{4}}
\multiput(-30,30)(60,0){2}{\circle{8}}
\multiput(-30,-30)(60,0){2}{\circle*{4}}
\multiput(-30,-30)(60,0){2}{\circle{8}}
\put(-1,0){\circle*{4}}
\put(0,-2){\bf x}
\put(0,60){\circle*{4}}
\put(0,-60){\circle*{4}}
\multiput(-56,0)(60,0){2}{\line(1,0){52}}
\multiput(-26,30)(0,-60){2}{\line(1,0){52}}
\multiput(-57,3)(30,30){2}{\line(1,1){24}}
\multiput(-57,-3)(30,-30){2}{\line(1,-1){24}}
\multiput(57,3)(-30,30){2}{\line(-1,1){24}}
\multiput(57,-3)(-30,-30){2}{\line(-1,-1){24}}
\put(3,3){\line(1,1){24}}
\put(3,-3){\line(1,-1){24}}
\put(-3,-3){\line(-1,-1){24}}
\put(-3,3){\line(-1,1){24}}
\put(-46,30){$\Phi^2$}
\put(38,30){$\Phi^1$}
\put(-46,-36){$\bar\Phi^1$}
\put(38,-36){$\bar\Phi^2$}
\put(4,58){$\Gamma$}
\put(-10,-66){$\bar\Gamma$}
\put(-12,3){$\Omega$}
\put(8,-9){$\Sigma$}
\put(86,20){
\begin{picture}(60,60)(-30,-30)
\put(0,0){\vector(1,1){20}}
\put(0,0){\vector(1,-1){20}}
\put(0,0){\vector(-1,-1){20}}
\put(0,0){\vector(-1,1){20}}
\put(24,24){$\cG^1$}
\put(24,-28){${\bar\cG}^2$}
\put(-28,-28){${\bar\cG}^1$}
\put(-28,24){$\cG^2$}
\end{picture}}
\put(86,-80){
\begin{picture}(60,80)(0,-60)
\put(0,2){\circle*{4}}
\put(-2,-20){\bf x}
\put(0,-36){\circle{8}}
\put(6,0){$Vir\times SU(2)$ primary fields}
\put(6,-20){Additional primary singlet}
\put(6,-40){Secondary fields appearing in the OPEs}
\end{picture}}
\put(232,-25){
\begin{picture}(70,70)(-35,-35)
\put(16,46){\it Short}
\multiput(-30,0)(60,0){2}{\circle*{4}}
\multiput(0,30)(0,-60){2}{\circle*{4}}
\put(-28,0){\line(1,0){56}}
\put(-29,1){\line(1,1){28}}
\put(-29,-1){\line(1,-1){28}}
\put(29,1){\line(-1,1){28}}
\put(29,-1){\line(-1,-1){28}}
\put(34,0){$\Psi^1$}
\put(-42,0){$\Psi^1$}
\put(4,28){$\Phi$}
\put(-10,-36){$\Pi$}
\end{picture}}
\put(170,45){
\begin{picture}(40,30)(0,0)
\put(0,20){\vector(1,0){30}}
\put(30,0){\vector(-1,0){30}}
\put(34,16){$A^+$}
\put(34,-4){$A^-$}
\end{picture}}
\end{picture}
\caption{\sl Long and short representations}
\end{figure}

\vskip 0.2cm\noindent
\underline{\sl Orbifold Conformal Field Theory of $\csg$}
\vskip 0.2cm
Now we consider the explicit construction of the orbifold conformal
field theory $\c^2/\Gamma$, starting from the $(4,4)$ theory of $\c^2$.
Let $X,\,\bar X$ and $Y,\,\bar Y$ be two complex bosonic fields,
$\psi_x,\,\bar\psi_x$ and
$\psi_y, \, \bar \psi_y$ two complex fermions.
They are normalized according to
\begin{eqnarray}
\partial X(z)\partial\bar X(w)&=&-{2\over (z-w)^2},\nonumber\\
\psi_x(z)\bar \psi_x(w)&=&-{2\over z-w}.
\end{eqnarray}
The N=4 superconformal algebra is realized by setting
\begin{eqnarray}
T(z)&=&-{1\over 2}(\partial X\partial \bar X+\partial Y\partial \bar Y)+
{1\over 4}(\bar \psi_x\partial\psi_x-\partial\bar\psi_x\psi_x+
\bar \psi_y\partial\psi_y-\partial\bar\psi_y\psi_y),\nonumber\\
A^i(z)&=&{1\over 4}\left[\matrix{i(\psi_x\psi_y+\bar\psi_x\bar\psi_y)\cr
\bar\psi_x\bar\psi_y-\psi_x\psi_y\cr
\psi_x\bar\psi_x+\psi_y\bar\psi_y}\right],\nonumber\\
\G^a(z)&=&{1\over \sqrt{2}}\left[\matrix{\bar\psi_x\cr i\psi_y}\right]
\partial X+
{1\over \sqrt{2}}
\left[\matrix{\bar\psi_y\cr -i\psi_x}\right]\partial Y,\nonumber\\
\bar\G^a(z)&=&{1\over \sqrt{2}}
\left[\matrix{\psi_x\cr -i\bar\psi_y}\right]\partial \bar X+
{1\over \sqrt{2}}
\left[\matrix{\psi_y\cr i\bar\psi_x}\right]\partial \bar Y.
\label{n=4flat}
\end{eqnarray}
The short representations can be easily obtained by looking at the
doublets that appear in the supercurrents. We have
\begin{equation}
\matrix{
\left(\Psi^a={1\over \sqrt{2}}
\left[\matrix{\psi_x\cr -i \bar\psi_y}\right],\Phi =-\partial X,
\Pi =-i\partial \bar Y\right),\cr
\left({1\over \sqrt{2}}\left[\matrix{\psi_y\cr i \bar\psi_x}\right],
-\partial Y,
i\partial \bar X\right)\,.}
\label{massless43}
\end{equation}
These representations satisfy the OPEs (\ref{masslessOPE2}).

The long representations are
\begin{eqnarray}
\Omega&=&{\rm exp}\,{i(p_x X+\bar p_x \bar X+p_y Y+\bar p_y \bar
Y)},\nonumber\\
\Phi^a&=&-i\sqrt{2}\left[\matrix{\bar p_x \bar\psi_x +\bar p_y \bar \psi_y\cr
i\bar p_x \psi_y-i \bar p_y \psi_x}\right]\,
\Omega,\nonumber\\
\bar \Phi^a&=&-i\sqrt{2}\left[\matrix{ p_x \psi_x +p_y \psi_y\cr
-i p_x \bar\psi_y+i  p_y \bar\psi_x}\right]\,
\Omega,\nonumber\\
\Gamma&=&2[\bar p_x \partial Y-\bar p_y \partial X+i(\bar p_x \bar \psi_x
+\bar p_y \bar \psi_y)(\bar p_x\psi_y-\bar p_y\psi_x)]\,
\Omega,\nonumber\\
\bar\Gamma&=&2[p_y \partial \bar X-p_x \partial \bar Y
+i(p_x \psi_x +p_y \psi_y)(p_y\bar\psi_x-p_x\bar \psi_y)]\,
\Omega,\nonumber\\
\Sigma&=&2[i(p_x\partial X-\bar p_x \partial \bar X+p_y \partial Y-\bar p_y
\partial \bar Y)
- (|p_x|^2-|p_y|^2)(\bar \psi_x\psi_x-\bar \psi_y\psi_y)\nonumber\\&&
-2 \bar p_x p_y\bar \psi_x \psi_y+2 p_x \bar p_y\psi_x \bar\psi_y]\,
\Omega.
\end{eqnarray}

Now we turn to the study of the orbifold conformal field theory,
and, after a brief review of the generalities
on orbifold constructions \cite{dixonharveywitten}, we will focus
on the case $\Gamma=A_{n-1}$.\par
The construction of the orbifold conformal field theory $\c^2/\Gamma$
begins with a Hilbert space projection onto $\Gamma$ invariant states. This
projection
can be represented
in lagrangian form as the sum over contributions of fields twisted in temporal
direction by all the elements of the group, i.e. $x(\sigma,\tau+2\pi)=
\gamma x(\sigma,\tau)$. n
hamiltonian language the twisted boundary conditions correspond to insertion
of the operator implementing $\gamma$
 in the Hilbert space, as it wiil be explained with more
details in section \ref{part}, and hence the sum $\sum_{\gamma\in \Gamma}g$
realizes the
projection operator onto $\Gamma$ invariant states. To obtain a modular
invariant theory we are forced to consider also twisted boundary conditions
in the spatial direction, i.e. $x(\sigma+2\pi,\tau)=\gamma x(\sigma,\tau)$;
from the stringy point of
view, these sectors correspond to the case in which the string is closed
only modulo a transformation of the group $\Gamma$. One may think to have a
different boundary condition for every element of the group; actually there
is a boundary condition for each conjugacy class of the group, for,
if the field obeys
\begin{equation}
x(z+1) = \gamma x(z)
\end{equation}
it also obeys
\begin{equation}
\eta x(z+1) = (\eta\gamma\eta^{-1})\eta x(z)
\end{equation}
where $\eta$ is any other element of $\Gamma$. So the sectors twisted by
$\eta\gamma\eta^{-1}$ are in fact all the same sector.\par
We have to introduce ``twist'' operators which applied to the vacuum
realize the change of sector in the Hilbert space, modifying the monodromy
properties of the fields. Such situation recalls what happens for
fermions, where we are explicitly able to construct
spin fields which change the boundary
conditions of the fermionic fields.\par
For the description of the monodromy properties of the fermions
$\psi_x(z)$, $\psi_y(z)$, $\tilde\psi_x(\bar z)$, $\tilde\psi_y(\bar z)$,
in the $A_{n-1}$ case,
namely
\begin{equation}
\matrix{
\psi_x({\rm e}^{2\pi i}z)={\rm e}^{2\pi i {k \over n}}\psi_x(z),&
\psi_y({\rm e}^{2\pi i}z)={\rm e}^{2\pi i {n-k \over n}}\psi_y(z),\cr
\tilde \psi_x({\rm e}^{-2\pi i}\bar z)={\rm e}^{2\pi i {n-k \over n}}
\tilde \psi_x(\bar z),&
\tilde \psi_y({\rm e}^{-2\pi i}\bar z)=
{\rm e}^{2\pi i {k \over  n}}\tilde \psi_y(\bar z),
}
\end{equation}
we introduce the
spin fields $s_x^{(k)}(z)$, $s_y^{(k)}(z)$ and their world-sheet
complex conjugates $\tilde s_x^{(k)}(\bar z)$, $\tilde s_y^{(k)}(\bar z)$.
 Their OPEs with the fermions are
\begin{eqnarray}
\psi_x(z)s^{(k)}_x(w)&=&(z-w)^{k\over n}{t^\prime}^{(k)}_x(w),\nonumber\\
\bar\psi_x(z)s^{(k)}_x(w)&=&{1\over (z-w)^{k\over n}}t^{(k)}_x(w)
\end{eqnarray}
and similar for the world-sheet complex conjugates.
Analogous formul\ae\ will hold for the fermions associated with
the $Y$ coordinate.
World-sheet complex conjugation means ($z\leftrightarrow
\bar z$, $h\leftrightarrow \bar h$). One has $\tilde\sigma_x^{(k)}=
\sigma_x^{(k)}$.

The spin fields can be represented by means of a bosonization:
\begin{eqnarray}
\matrix{
\psi_x=-i\sqrt{2}\,{\rm e}^{iH_x},&&\bar\psi_x=-i\sqrt{2}\,{\rm e}^{-iH_x},\cr
&s^{(k)}_x={\rm e}^{i{k\over n}H_x},&\cr
{t^\prime}^{(k)}_x=-i\sqrt{2}\,{\rm e}^{i\left(1+{k\over n}\right)H_x},&&
t^{(k)}_x=-i\sqrt{2}\,{\rm e}^{-i\left(1-{k\over n}\right)H_x}.}
\label{spintwist}
\end{eqnarray}

The twist operators for the bosonic fields were introduced in
\cite{dixon} and they are denoted by
$\sigma_x^{(k)}(z,\bar z)$ and $\sigma_y^{(k)}(z,\bar z)$,
$k=1,\ldots n$.
In a neighborhood of a twist field located at the origin the fields $X$
and $Y$ have the monodromy properties
\begin{eqnarray}
\matrix{
X({\rm e}^{2\pi i}z,{\rm e}^{-2\pi i}\bar z)={\rm e}^{2\pi i {k \over n}}
X(z,\bar z),&
Y({\rm e}^{2\pi i}z,{\rm e}^{-2\pi i}\bar z)={\rm e}^{2\pi i {n-k \over  n}}
Y(z,\bar z).}
\end{eqnarray}
Correspondingly,
the OPEs of the twist fields
with $\partial X(z)$, $\partial \bar X(z)$, $\bar \partial X(\bar z)$
and $\bar \partial\bar X(\bar z)$ are
\begin{eqnarray}
\partial X(z)\sigma_x^{(k)}(w,\bar w)&=&
{1\over (z-w)^{1-{k\over n}}}\tau^{(k)}_x(w,\bar w),
\nonumber\\
\partial \bar X(z)\sigma_x^{(k)}(w,\bar w)&=&{1\over (z-w)^{k\over n}}
{\tau^\prime}^{(k)}_x(w,\bar w),\nonumber\\
\bar\partial X(\bar z)\sigma_x^{(k)}(w,\bar w)&=&
{1\over (\bar z-\bar w)^{k\over n}}{\tilde\tau}^{\prime(k)}_x(w,\bar w),
\nonumber\\
\bar\partial \bar X(\bar z)\sigma_x^{(k)}(w,\bar w)&=&{1\over (\bar z-
\bar w)^{1-{k\over n}}}
{\tilde\tau}^{(k)}_x(w,\bar w).
\end{eqnarray}
The $\tau$-fields are called {\it excited twist fields}.
Similar formul\ae\ hold for the $Y$ coordinate and the corresponding twist
fields. Unfortunately, we don't have an
explicit construction  of the
bosonic twists in terms of the fundamental bosonic
fields, a fact which makes more difficult the computation of correlation
functions and fusion rules involving twist fields \cite{dixon}.\par
The operator content of the orbifold conformal field theory is given by
$\Gamma$
invariant operators (coming from the untwisted sector) and product of twist
fields and $\Gamma$ invariant operators (from the twisted sectors). From
the operatorial point of view, the projection onto invariant states is needed
to obtain a set of mutually local operators.\par
The computation of the expectation value of the stress-energy tensor
in the presence of twist fields \cite{dixon} gives the conformal
dimension of the twist $h_{\sigma^{(k)}}={1\over 2}(k/n)(1-k/n)$. From the
bosonization rules (\ref{spintwist}) we learn more directly the conformal
dimension of the
spin field $h_{s^{(k)}}={1\over 2}(k/n)^2$.\par

{}From the bosonization rules, we have
\begin{eqnarray}
\psi_x(z)t^{(k)}_x(w)&=&-{2\over (z-w)^{1-{k\over n}}}s^{(k)}_x(w),\nonumber\\
\bar\psi_x(z){t^\prime}^{(k)}_x(w)&=&-{2\over (z-w)^{1+{k\over
n}}}s^{(k)}_x(w).
\end{eqnarray}
The other OPEs of this kind
(i.e. for $\psi_x(z){t^\prime}^{(k)}_x(w)$ and $\bar\psi_x(z)t^{(k)}_x(w)$)
are regular. The analogous formul\ae\
for the OPEs between bosons and excited twist fields will be derived later on,
when studying systematically the product of representations of the orbifold
conformal field theory.
Moreover, we have, for $k+k^\prime<n$
\begin{eqnarray}
s_x^{(k)}(z)s_x^{(k^\prime)}(w)&=&(z-w)^{{k\over n}{k^\prime\over n}}s_x^
{(k+k^\prime)}(w),
\nonumber\\
s_x^{(n-k)}(z)s_x^{(n-k^\prime)}(w)&=&
{i\over \sqrt{2}}
(z-w)^{{k\over n}{k^\prime\over n}}\psi_x(z)s_x^{(n-k-k^\prime)}(w).
\label{tut}
\end{eqnarray}
These formul\ae\ will be useful in the following.

As for the OPEs between twist fields, it is reasonable to assume
\begin{eqnarray}
\sigma_x^{(k)}(z,\bar z)\sigma_x^{(k^\prime)}(w,\bar w)&=&{C_{k,k^\prime}^
{k+k^\prime}\over
|z-w|^{2{k\over n}{k^\prime\over n}}}
\sigma_x^{(k+k^\prime)}(w,\bar w)
\hskip 8pt {\rm for}\, \, k+k^\prime<n,\nonumber\\
\sigma_x^{(k)}(z,\bar z)\sigma_x^{(k^\prime)}(w,\bar w)&=&{C_{k,k^\prime}^
{k+k^\prime-n}\over
|z-w|^{2\left(1-{k\over n}\right)\left(1-{k^\prime\over n}\right)}}
\sigma_x^{(k+k^\prime-n)}(w,\bar w)
\hskip 8pt {\rm for}\, \, k+k^\prime>n,\nonumber\\
\label{tutu}
\end{eqnarray}
where $C_{k,k^\prime}^{k+k^\prime}$ and $C_{k,k^\prime}^{k+k^\prime-n}$
are certain coefficients (a sort of structure constants)
that we do not need to specify here.

We now study the representations of the orbifold theory. The representations
that are defined by means of twist and spin fields and not only
with the fields of the ${\bf C}^2$-theory will be called
{\it twisted representations}.
The twisted short representations mix the left and right sectors. The
lowest component of the short representations are
\begin{equation}
\Psi_k\sp{\um,\um}{\um,\um}^{a\tilde b}
(z,\bar z)=\sigma^{(k)}_x(z,\bar z) \sigma^{(n-k)}_y(z,\bar z)
\sp{s_x^{(k)}s_y^{(n-k)}}{{i\over 2} t_x^{(k)}t_y^{(n-k)}}^a(z)
\sp{\tilde s_x^{(k)}\tilde s_y^{(n-k)}}
{{i\over 2} \tilde t_x^{(k)}\tilde t_y^{(n-k)}}^{\tilde b}(\bar z).
\label{twistshort}
\end{equation}
The field content of the representation will be denoted by
\begin{equation}
\Big(\Psi_k^{a\tilde a},(\Psi\tilde\Phi)^a_k,
(\Psi\tilde\Pi)_k^a,(\Phi\tilde\Psi)_k^{\tilde a},
(\Pi\tilde\Psi)_k^{\tilde a},
(\Phi\tilde\Phi)_k,
(\Phi\tilde\Pi)_k,(\Pi\tilde\Phi)_k,
(\Pi\tilde\Pi)_k\Big).
\label{leftright}
\end{equation}
The notation of the fields is reminiscent of the fact that they
transform as the tensor product of two short representations
[see eq.(\ref{masslessOPE2})],
one in the left sector and
one in the right sector. In this spirit we could have written
$(\Psi\tilde\Psi)_k^{a\tilde a}$ instead of $\Psi_k^{a\tilde a}$.
However, we note that the twisted representations are not such a tensor
product, due to the fact that the twist fields depend both on $z$ and
$\bar z$. The operators that are needed for the description of the
deformations of the conformal field theory are
$(\Phi\tilde\Phi)_k$, $(\Phi\tilde\Pi)_k$, $(\Pi\tilde\Phi)_k$ and
$(\Pi\tilde\Pi)_k$.
We have to work out the required OPEs in order to get their expressions.
To do this we write
\begin{eqnarray}
\partial X(z){\tilde\tau}^{\prime (k)}_x(w,\bar w)&=&
{1\over (z-w)^{1-{k\over n}}}\Delta{\tilde\tau}^{\prime(k)}_x(w,\bar w),
\nonumber\\
\partial \bar X(z)\tilde\tau_x^{(k)}(w,\bar w)&=&{1\over (z-w)^{k\over n}}
\bar\Delta\tilde\tau^{(k)}_x(w,\bar w),
\nonumber\\
\bar\partial X(\bar z){\tau^\prime}_x^{(k)}(w,\bar w)&=&
{1\over (\bar z-\bar w)^{k\over n}}\tilde{\bar\Delta}{\tau^\prime}
^{(k)}_x(w,\bar w),
\nonumber\\
\bar\partial \bar X(\bar z)\tau_x^{(k)}(w,\bar w)&=&{1\over (\bar z-
\bar w)^{1-{k\over n}}}
\tilde\Delta\tau^{(k)}_x(w,\bar w).
\end{eqnarray}
The $\Delta\tau$-fields are doubly-excited twist fields.
We give the explicit expressions of the fields that describe
the deformations of the conformal field theory, i.e.
$(\Phi\tilde\Phi)_k$, $(\Phi\tilde\Pi)_k$, $(\Pi\tilde\Phi)_k$ and
$(\Pi\tilde\Pi)_k$.
Omitting the superscripts $k$ and $n-k$ in the
$X$-fields and $Y$-fields, respectively, they are
\begin{eqnarray}
(\Phi\tilde\Phi)_k&=&{1\over 2}[\tilde\Delta\tau_x\sigma_y
t_x s_y \tilde t_x\tilde s_y -
\tilde\tau_x\tau_y s_x t_y \tilde t_x
\tilde s_y
-\tau_x\tilde\tau_y t_x s_y \tilde s_x
\tilde t_y +\sigma_x\tilde\Delta\tau_y s_x
t_y \tilde s_x\tilde t_y ],
\nonumber\\
(\Phi\tilde\Pi)_k&=&{i\over 2}[\Delta{{\tilde\tau}^\prime}_x
\sigma_y t_x s_y \tilde s_x\tilde t_y +
\tau_x{{\tilde\tau}^\prime}_y t_x s_y \tilde
t_x\tilde s_y
-{{\tilde\tau}^\prime}_x\tau_y s_x t_y \tilde
s_x\tilde t_y -\sigma_x\Delta{{\tilde\tau}^\prime}_y
s_x t_y \tilde t_x\tilde s_y ]
\nonumber\\
(\Pi\tilde\Phi)_k&=&{i\over 2}[\bar\Delta\tilde\tau_x\sigma_y
s_x t_y \tilde t_x\tilde s_y -
{\tau^\prime}_x\tilde\tau_y s_x t_y
\tilde s_x\tilde t_y
+\tilde\tau_x{\tau^\prime}_y t_x s_y \tilde
t_x\tilde s_y -\sigma_x\bar \Delta\tilde\tau_y
t_x s_y \tilde s_x\tilde t_y ],
\nonumber\\
(\Pi\tilde\Pi)_k&=&-{1\over 2}[\tilde{\bar\Delta}{\tau^\prime}
_x\sigma_y
s_xt_y \tilde s_x\tilde t_y +
{{\tilde\tau}^\prime}_x{\tau^\prime}_y t_x s_y
\tilde s_x\tilde t_y
+{\tau^\prime}_x{{\tilde\tau}^\prime}_y s_x t_y
\tilde t_x \tilde s_y +\sigma_x
\tilde{\bar \Delta}{\tau^\prime}_y t_x s_y \tilde
t_x \tilde s_y ].\nonumber\\
\end{eqnarray}
Consistency, i.e. the fact that the same fields of the short representation
(\ref{leftright})
can be reached from different paths when one repeatedly applies the
supercurrents, implies
\begin{eqnarray}
\partial X(z)\tilde\tau_x^{(k)}(w,\bar w)&=&
{1\over (z-w)^{1-{k\over n}}}\tilde\Delta\tau^{(k)}_x(w,\bar w),
\nonumber\\
\partial \bar X(z){\tilde\tau}^{\prime (k)}_x(w,\bar w)
&=&{1\over (z-w)^{k\over n}}
\tilde{\bar\Delta}{\tau^\prime}^{(k)}_x(w,\bar w),
\nonumber\\
\bar\partial X(\bar z)\tau_x^{(k)}(w,\bar w)&=&
{1\over (\bar z-\bar w)^{k\over n}}
\Delta{\tilde\tau}^{\prime (k)}_x(w,\bar w),
\nonumber\\
\bar\partial \bar X(\bar z){\tau^\prime}_x^{(k)}(w,\bar w)&=&{1\over (\bar z-
\bar w)^{1-{k\over n}}}
\bar\Delta{\tilde\tau}^{(k)}_x(w,\bar w).
\end{eqnarray}

Eventual variants of the above representation (\ref{leftright}),
obtained by substituting in $\Psi_k^{1\tilde1}$
the product
of twist fields
$s_x^{(k)}s_y^{(n-k)}$$\tilde s_x^{(k)}\tilde s_y^{(n-k)}$ with
the product
$s_x^{(n-k)}s_y^{(k)}$ $\tilde s_x^{(k)}\tilde s_y^{(n-k)}$
or $s_x^{(n-k)}s_y^{(k)}$ $\tilde s_x^{(n-k)}\tilde s_y^{(k)}$
or $s_x^{(k)}s_y^{(n-k)}$ $\tilde s_x^{(n-k)}\tilde s_y^{(k)}$ and
making the analogue for the $t$-fields, surely satisfy the correct OPEs.
However they are not good representations of the orbifold theory, for
sectors with $k\neq 0$. In fact, by definition of twisted sectors,
the $X$-spin fields and $X$-twist fields must carry the same superscript,
say $k$; in this case the $Y$-spin fields and $Y$-twist fields must
carry the superscript $n-k$.
In conclusion, the short representations of the orbifold conformal
field theory are four (those of the untwisted sector)
plus one for each twisted sector.

In this way we recover at the level of conformal field theory the correct
counting of moduli parameters. Comparing with the abstract Hodge diamond
[see eq. (\ref{introdhodgediamond})]
we see that $h^{(1,0)}=0$ and $h^{(1,1)}=4+\tau$
($|\tau|=n-1$). Indeed $h^{(1,0)}=0$ is explained by the fact that the
untwisted short representations
$\Psi\sp{\um,0}{\um,0}^a={1\over \sqrt{2}}\left[\matrix{\bar
\psi_x\cr i\psi_y}\right]$ or ${\Psi^\prime}\sp{\um,0}{\um,0}^a=
{1\over \sqrt{2}}\left[\matrix{\bar
\psi_y\cr -i\psi_x}\right]$ of eq. (\ref{massless43}) that are present
in the $\c^2$ case, are deleted by the projection onto $\Gamma$-invariant
states in the $\c^2/\Gamma$ theory. On the other hand, $h^{(1,1)}=4+\tau$
is explained by the fact that the orbifold theory contains the $|\tau|$
twisted short
representations in addition to the untwisted ones. The untwisted
representations correspond to the $1+3$ non-normalizable $(1,1)$-forms
that have to be deleted in compact support cohomology.
The abstract Hodge diamond is thus
\begin{equation}
\matrix{
&&1&&\cr
&0&&0&\cr
1&&4+|\tau |&&1\cr
&0&&0&\cr
&&1&&}.
\label{alehd}
\end{equation}

Finally, the
$(4,4)$-theory corresponding to the smooth manifolds ${\cal M}_\zeta$
is obtained by perturbing the $\c^2/\Gamma$ theory with the operator
\begin{equation}
{\cal O}={\rm exp}\left\{ \sum_{k=1}^{\tau}
\int d^2 z [\xi_k^1\,(\Phi\tilde\Phi)_k+\xi_k^2\,(\Phi\tilde\Pi)_k+\xi_k^3\,
(\Pi\tilde\Phi)_k+\xi_k^4\,(\Pi\tilde\Pi)_k]\right\},
\label{orbifolddeformation}
\end{equation}
where the $4 \times |\tau|$ parameters $\xi_k^i$ ($i=1,2,3,4$),
that can be arranged into a quaternion for each value of $k$,
describe the parameters of the K\"ahler class, complex structure
and torsion deformations. In the geometric treatment we have so far
considered only the HyperK\"ahler deformations ($3\times |\tau|$
parameters), however the conformal field theory contains also the deformations
of the axion tensor $B_{\mu\nu}$ leading to the torsion deformations.
The problem of identifying the moduli $t_n$ of (\ref{grouptheory21}) in terms
of the moduli $\xi$ of (\ref{orbifolddeformation}) remains open.
So far, we have introduced various coordinate systems in the moduli-space:
the $\xi$ coordinates appearing in (\ref{orbifolddeformation}), that
are a sort of ``flat coordinates'' and that describe all possible deformations,
the $\zeta$-coordinates of the momentum map approach (see
section \ref{construction}) that parametrize the deformations of the
complex structure and K\"ahler class
and the $t$-coordinates that parametrize the chiral ring
${\cal R}=\c^2[x,y,z]/\partial W$. Formula (\ref{def2}) established
the relation between the $t$ and the $\zeta_+$ parameters
[called $\lambda$ in the context of formula (\ref{def2})].
The extension of this identification to the $\xi$ parameters
is an open problem.

We conclude this section by studying the operator
product of two short representations, that is
the set of all the OPEs between the fields of the
two multiplets.
This operation is interesting, because it
provides examples of twisted long representations.
For simplicity, we concentrate only on the left part of the representations,
thus omitting all tilded fields. Let us consider the product of an untwisted
representation [say the one of (\ref{massless43})] with the representation
(\ref{leftright}). The untilded part of $\Psi^{a\tilde b}_k$ will be denoted
by $\Psi_k^a$. The two singlets of the untilded part of representation
(\ref{leftright}) will be denoted by $\Phi_k$ and $\Pi_k$, so that the notation
for the entire representation will be $(\Psi^a_k,\Phi_k,\Pi_k)$.
The product in question gives a long representation in which
the lowest weight primary
field $\Omega_k$ is
\begin{equation}
\Omega_k={1\over \sqrt{2}}\sigma_x^{(k)}\sigma_y^{(n-k)}t_x^{(k)}s_y^{(n-k)},
\label{we}
\end{equation}
and its weight is $h=1-k/n$.
In particular, we have
\begin{equation}
\Psi^a(z)\Psi^b_k(w,\bar w)=\delta^{ab}{1\over (z-w)^{k\over n}}
\Omega_k(w,\bar w).
\end{equation}
One can check that
consistency with the general OPEs of (\ref{massivegen}) fixes
the OPEs between $\partial X$, $\partial \bar X$ and the excited twist
fields. We only give some examples
\begin{eqnarray}
\partial X(z)\tau_x^{(k)}(w,\bar w)&=&
{\Delta \tau_x^{(k)}(w,\bar w)\over (z-w)^{1-{k\over n}}},
\nonumber\\
\partial \bar X(z)\tau_x^{(k)}(w,\bar w)&=&
-2{k\over n}{1\over (z-w)^{1+{k\over n}}}
\sigma_x^{(k)}(w,\bar w)+
{\bar\Delta \tau_x^{(k)}(w,\bar w)\over (z-w)^{k\over n}},
\nonumber\\
\partial X(z){\tau^\prime}_x^{(k)}(w,\bar w)&=&-2\left(1-{k\over n}\right)
{1\over (z-w)^{2-{k\over n}}}\sigma_x^{(k)}(w,\bar w)+
{\Delta {\tau^\prime}_x^{(k)}(w,\bar w)\over (z-w)^{1-{k\over n}}},
\nonumber\\
\partial \bar X(z){\tau^\prime}_x^{(k)}(w,\bar w)&=&
{\bar\Delta {\tau^\prime}_x^{(k)}(w,\bar w)\over (z-w)^{k\over n}},
\end{eqnarray}
for certain $\Delta \tau_x^{(k)}(w,\bar w)$,
$\bar\Delta \tau_x^{(k)}(w,\bar w)$,
$\Delta {\tau^\prime}_x^{(k)}(w,\bar w)$
and $\bar\Delta {\tau^\prime}_x^{(k)}(w,\bar w)$.
Going on in this way, OPEs involving doubly excited twist fields can also
be found.

We now consider the operator product of two short representations
of the twisted sectors. Using (\ref{tut}) and (\ref{tutu}),
we find that this product gives one of the twisted
long representations that we have just found.
So, the
product of two twisted short representations of the orbifold is also
equal to the
product of an untwisted short representation and
a third twisted short representation of the orbifold. Precisely,
\begin{equation}
{R}_k(z,\bar z){R}_{k^\prime}(w,\bar w)=
{R}_0^{\epsilon_{k+k^\prime}}(z,\bar z)
{R}_{k+k^\prime \,\, {\rm mod} \, n}(w,\bar w),
\end{equation}
where ${R}_k(z,\bar z)$ denotes in compact form the twisted massless
representation of the $k^{th}$-sector, while
${R}_0^{\epsilon_{k+k^\prime}}(z,\bar z)$ is one of the two
untwisted short representations, which of them depending on the sign
of $k+k^\prime-n$.

In this way we can define a natural
product between the short representations of the orbifold,
that still gives a short representation of the orbifold and
that satisfies the cyclic property of the $A_{n-1}$ subgroup of $SU(2)$,
namely
\begin{equation}
{R}_k\otimes {R}_{k^\prime}={R}_{k+k^\prime\,\,{\rm mod}\, n}.
\end{equation}

The last remark we make concerns the chiral ring \cite{lercheetal} of the
conformal field theory under consideration.
Let
\begin{equation}
{\cal O}_k=\sigma_x^{(k)}\sigma_y^{(n-k)}s_x^{(k)}s_y^{(n-k)}\tilde s_x^{(k)}
\tilde s_y^{(n-k)}
\label{twist}
\end{equation}
be the operator that, acting on the vacuum, gives the vacuum of the $k^{th}$
twisted sector. Viewing the N=4 theory as an N=2 theory, the ${\cal O}_k$ are
chiral operators. All of them have charge 1 (this is
the $U(1)$ charge corresponding to the current $A_3 (z)$).
There must also exist a unique
chiral operator of charge ${c\over 3}=2$ and conformal weight $1$.
This operator is $A^+$. These operators $(A^+, \cO_k )$ together with
the identity span the chiral ring $\cC$ of the $N=2$ theory, which
happens to have only integral $U(1)$ charges.
As a matter of fact, one verifies that
\begin{eqnarray}
{\cal O}_k{\cal O}_{k^\prime}&=&0 \hskip 2truecm
 {\rm if}\, k+k^\prime \neq n,\nonumber\\
{\cal O}_k{\cal O}_{n-k}&\sim&A^+.
\label{multtable}
\end{eqnarray}
It is important to stress that this N=2 chiral ring is not the ring ${\cal R}
=C[x,y,z]/\partial W$, although the short representations are in one-to-one
correspondence with the elements of ${\cal R}$. It remains a so far unsolved
problem to interpret the multiplication rule of ${\cal R}$ within
the conformal field theory.
In particular, we can perform the standard topological twist; the chiral
ring $\cC$ coincides then with the set of physical fields of the
topological model. The multiplication table (\ref{multtable}) of the
unperturbed chiral ring $\cC$ yields the two point function (the
topological metric) which is just constant in parameter space. One can
also perturb the topological model by inserting into the correlators the
exponential of the integrated physical fields, namely the second
components of the $\cO_k$ chiral multiplets. This is nothing else but
the operator $\cO$ in eq.(\ref{orbifolddeformation}). As it is well
known the parameters $\xi$ of the perturbation of a topological model
around its conformal point are the {\it flat coordinates} (see
\cite{dvv} whose relation with the parameters $t$ in the deformed
Landau-ginzburg potential (if one exists) is given by the solution of a
complicated but well-defined uniformization problem. Therefore if the chiral
ring $\cC$ of the $N=2$ were identified with the chiral ring $\cR$ then
the relation between the $x_i$ parameters in
eq.(\ref{orbifolddeformation}) and the $t$-parameters in eq.s
(\ref{grouptheory21},\ref{def2}) could be retrieved using the methods
developed for Topological Landau-Ginzburg models. This shows that
finding the relation between $\xi$ and $t$ or finding the relation
between the two rings $\cC$ and $\cR$ or finding the relation between
the $\sigma$-model on an ALE space and a suitable Landau-Ginzburg model
are just three different aspects of the same problem.

\section{Partition function and $N=4$ characters}
\label{part}

As stressed several times the moduli of a $(4,4)$ theory are the
highest components of short representations.
The characters of the short representations were computed by Eguchi and
Taormina in
\cite{eguchi2}. They read
\footnote{in the following we use for the theta-functions with
characteristic two equivalent notations:
$\theta{0\brack 0}\equiv \theta_3$,
$\theta{1\brack 0}\equiv \theta_2$,
$\theta{0\brack 1}\equiv \theta_4$,
$\theta{1\brack 1}\equiv \theta_1$.}:
\begin{eqnarray}
&&ch^{NS}_0(l=0;\tau ;z)=2\left( {\theta_1(z)\over \theta_3(0)}\right)^2
+\left( {q^{-{1\over 8}}\over \eta(\tau)} - 2h_3(\tau )\right)\left(
{\theta_3(z)\over \eta(\tau )}\right)^2\nonumber\\
&&ch^{NS}_0(l={1\over 2};\tau ;z)= -\left({\theta_1(z)\over \theta_3(0)}
\right)^2 + h_3(\tau )\left({\theta_3(z)\over \eta(\tau )}\right)^2
\end{eqnarray}
where $h_3$ is defined by
\begin{equation}
h_3(\tau )= {1\over \eta(\tau )\theta_3(0)}\sum_m{q^{{1\over 2}m^2-{1\over 8}}
\over 1+q^{m-{1\over 2}}}
\end{equation}
The characters of the long representations are instead given by:
\begin{equation}
ch^{NS}(h;\tau ;z)={q^{h-{1\over 8}}\over \eta(\tau )}\left({\theta_3(z)\over
\eta(\tau )}\right)^2
\end{equation}
\par
The above formulae apply to a general N=4 theory. We are interested in the
specific case of the orbifold theories discussed in the previous section.
We want to explore their spectrum
by investigating their partition function and by
decomposing it into characters of the N=4 algebra.

Consider the partition function for the $
A_{k-1}$models;
we have to sum contributions from the twisted sector, i.e. to sum over
different boundary conditions in the spatial direction. To obtain a modular
invariant partition function, we have to twist also in the time direction.
As we have already explained in section \ref{cft}, the twisted
boundary conditions
are associated with the non-trivial conjugacy class of the group.
In the abelian case $A_{k-1}$,
we have a conjugacy class for every element of the cyclic group;
therefore the boundary conditions are parametrized by an integer $0\leq i\leq
k-1$. If we denote $\nu = e^{2\pi i/k}$, the building blocks for the partition
function are
\begin{equation}
Z_{r,s} = (q\bar q)^{-c/24} \tr_{\nu^r}\ \hat \nu^s q^{L-0}\bar q^{\bar L_0}
e^{2\pi izJ_3+2\pi i\bar zJ_3}\, ,
\end{equation}
where $\hat \nu$ is the operator on the Hilbert space which implements the
action of the generator of $A_{k-1}$. $Z_{r,s}$ is the
partition function twisted
by $\nu^r$ in the spatial direction and by $\nu^s$ in the temporal one.\par
A modular invariant  partition function can be constructed by summing over all
boundary conditions
\begin {equation}
Z = {1\over k}\sum_{r,s} Z_{r,s}C_{r,s}
\label{orbpart}
\end{equation}
with coefficients $C_{r,s}=1$.
In terms of Hilbert space states, the sum over the spatial
conditions takes into account the existence of several sectors of the Hilbert
space, while the sum over temporal conditions, at fixed spatial ones,
realizes the necessary projection of the theory onto group invariant states.
\par
We have to sum over the twisted boundary conditions for bosons and fermions
dictated by the orbifold construction and, independently, over the four
spin structures of fermions to take into account the existence of
Ramond (R) and Neveu-Schwarz (NS)
sectors. To compute the total partition function \cite{ginsparg} we begin with
the contribution from the untwisted sector. Decomposing the fields $X,Y,
\psi_x,\psi_y$ in Fourier modes $(\alpha,\beta,\lambda,\mu )$
with the
commutation
relations
\begin{eqnarray}
&&[\bar \alpha_n,\alpha_m]=n\delta_{n+m,0}\nonumber\\
&&\{\bar\lambda_n,\lambda_m\}=\delta_{n+m,0}
\end{eqnarray}
and similar for $\beta,\mu$, we implement the group transformation on the
Hilbert space via
\begin{equation}
(\alpha_n,\bar\beta_n,\lambda_n,\bar\mu_n ) \rightarrow e^{2\pi i/k}
(\alpha_n,\bar\beta_n,\lambda_n,\bar\mu_n )
\end{equation}
In the untwisted sector we take a basis of eigenvectors of $L_0,\alpha_0,
\beta_0$ to perform the trace on Hilbert space; note that in this sector the
zero modes of bosonic fields (the momentum) commute with $L_0$. The Fock
space is constructed by applying the raising operators to the vacuum and to
the eigenvectors of the momentum $|p_x,p_y\rangle$. The trace with $\hat
\nu$ inserted picks out contributions only from the vacuum, because $\hat
\nu$ is not diagonal on the momentum eigenvectors. The computation
is now straightforward \cite{ginsparg} and reduces to the computation of
the partition function of free bosons and fermions with twisted boundary
conditions. The contribution of the unprojected trace is the partition function
of the flat N=4 space
\begin{equation}
{1\over k} {1\over (\eta\bar\eta )^4}\int dp_xdp_y \ q^{p_x\bar p_x +
p_y\bar p_y}\
{1\over 2}\sum \left|{\theta {a\brack b}(z)\over \eta}\right|^4\,\, .
\end {equation}
In the sector r-times twisted in the temporal direction, the standard $\eta$
function of a boson (for example) is replaced by the following
infinite product
\begin{equation}
q^{-{1\over 24}}{1\over \Pi_{n=1}^{\infty}(1 - \nu^r q^n)}\,\, .
\end{equation}
In the twisted sector of the theory the Hilbert space is constructed by
the application of oscillators to the twisted vacua,
obtained by applying the twist fields to the true vacuum. This explains a
further factor of $q^{1/2}$ to the energy due to
the conformal dimension of the twist fields $O_i$ (see eq. \ref{twist}).
 The oscillators
have now fractional indices and so they contribute fractional powers of $q$
in the infinite product which replaces the $\eta$ function,
exactly as it happens
to free fermions when we change the spin structure (i.e. boundary conditions).
{}From
\begin{equation}
i\partial X = \sum_n {\alpha_n\over z^{n+1}},\ \ \  i\psi_x = \sum_n {\lambda_n
\over z^{n+1/2}}\  (NS), ...
\end{equation}
and the transformation $\partial X \rightarrow e^{2\pi is/k}\partial X$,
... under the group $A_{k-1}$
we learn the modings
\begin{eqnarray}
&&\alpha_n,\bar\beta_n,\lambda_n,\bar\mu_n \ (P) ,\ \ \ \ \ n
\epsilon Z-{s\over k}
\nonumber\\
&&\bar\alpha_n,\beta_n,\bar\lambda_n,\mu_n \ (P) ,\ \ \ \ \
n\epsilon Z+{s\over k}\,\, .\end{eqnarray}
Collecting all these informations the general
contribution from antiperiodic-antiperiodic fermions (to give an explicit
example) is
\begin{eqnarray}
&&q^{-2/24}\prod_{n=0}^{\infty}\,(1 + e^{2\pi iz}\nu^r q^{n+{s\over k}
+{1\over 2}})(1+e^{2\pi iz}\bar\nu^rq^{n+{k-s\over k}-{1\over 2}})
\times \nonumber\\
&&\hskip 20pt\times(1 + e^{-2\pi iz}\nu^r q^{n+{s\over k}+{1\over 2}})
(1+e^{-2\pi iz}\bar\nu^rq^{n+{k-s\over k}-{1\over 2}})
\end{eqnarray}
from the triple product identity \cite{ginsparg}
\begin{equation}
\prod_{n=0}^{\infty}(1-q^{n+1})(1+q^{n+1/2}\omega )(1+q^{n+1/2}\omega^{-1}) =
\sum_{n\in Z} q^{n^2/2}\omega^n
\end{equation}
with $\omega = \nu^rq^{s/k}$
we obtain
\begin{eqnarray}
&&q^{1/24}(1 + e^{2\pi iz}\nu^r q^{n+{s\over k}+{1\over 2}})
(1+e^{2\pi iz}\bar\nu^rq^{n+{k-s\over k}-{1\over 2}})=\nonumber\\
&&=\o 1{\eta(q)}\sum_Z q^{n^2/2}\nu^{nr}e^{2\pi inz}q^{ns/k}=
{\theta_3(z+(r+s\tau )/k)\over \eta}.
\end{eqnarray}
The introduction of the fermionic  spin structures simply shifts some
parameters in the $\theta$ function. For simplicity, we collect only the
results for the NS sector. Fermions contribute two factors like the one above,
while for the same reasoning bosons (collecting with care also the $q^{1/2}$
factor in the twisted sector) contribute a factor ${\eta^2\over \theta_1}$.
The building blocks (in the NS sector) of the partition function are,
for a fixed fermionic spin structure, say ${0\brack 0}$,
\begin{equation}
Z_{r,s}{\scriptstyle{0\brack 0}} = {\theta_3(z+(r+s\tau )/k)
\theta_3(z-(r+s\tau )/k)\over
\theta_1((r+s\tau )/k)^2}\,\, .
\end{equation}
With the same technique we can compute the partition function for all
boundary conditions $Z{a\brack b}$ and from these the analogous of the
$i=(0,v,s,\bar s)$ characters for the flat space
and finally the heterotic partition function (see \cite{previous})
\begin{equation}
Z = \sum_{i,\bar \imath} Z_{i,\bar \imath}^{(9,9)}
Z_{i,\bar \imath}^{(6,6)}B_i^{(-2)}
\left (B_{\bar \imath}^{(E_8^{\prime}\times SO(6))}\right )^*\,\, .
\end{equation}
The next step to obtain the spectrum of the theory is to decompose the
partition
function into characters of the N=4 algebra.\par
If we are only interested in the field content of the orbifold conformal
field theory (ignoring internal dimensions), the
$Z {a\brack b}$ have to be summed with certain coefficients
to obtain a modular invariant
``partition function'' :
\begin{equation}
Z {\scriptstyle{a\brack b}} = Z {\scriptstyle{a\brack b}}_{\rm flat\, space} +
\sum_{r,s} \left|Z_{r,s}{\scriptstyle{a\brack b}}\right|^2\,\, .
\end{equation}
In the Eguchi-Hanson ($A-1$) case the above partition function (at $z=0$,
summed
over all boundary conditions) is easily
computed:
\begin{equation}
Z_{EH} = {1\over 4}{1\over (\eta\bar\eta)^4}\int dp_xdp_y \,\,
q^{p_x\bar p_x + p_y\bar p_y}
\sum_i\left|{\theta_i\over \eta}\right|^4 + {1\over 4}\sum_{i,j}
\left|{\theta_i\over\theta_j}\right|^4\,\, .
\end{equation}
It is modular invariant by inspection.\par
In the same spirit, the flat space contribution is
\begin{equation}
Z_{\rm flat\, space} = {1\over |Im\tau|^2|\eta |^8}\sum_i{1\over 2}\left|
{\theta_i\over\eta}\right|^4
\end{equation}
The momentum integral annihilates the contribution of short representation
which are a zero measure set with respect to the continuum spectrum.
We expect that the flat space partition function is an integral over the
continuum spectrum of the theory of long representations with characters
\begin {equation}
ch^{NS} = {q^{h-1/8}\over \eta}{\theta_3^2\over \eta^2}
\end{equation}
The continuum spectrum is realized by exponential fields or exponential
 fields
multiplied by combinations of derivatives of the bosonic fields and the
singlets
that we can realize with fermions such that the entire field is an N=4 primary.
A combinatorial computation shows that this sum reconstructs exactly
the three factors of $\eta$ needed to obtain $Z_{\rm flat\, space}$\par
The twisted sectors $Z_{s,r}$ do not receive contributions from
the continuum spectrum given by the exponentials. In the Eguchi-Hanson case
the explicit form of $Z_{r,s}$ at $z\neq 0$ is
\begin{eqnarray}
Z_{01}(z) &=& \left( {\theta_1(z)\over \theta_3}\right)^2 - \left( {\theta_2
\over \theta_4}\right)^2\left( {\theta_3(z)\over \theta_3}\right)^2
\nonumber\\
Z_{10}(z) &=& -\left( {\theta_1(z)\over \theta_3}\right)^2 - \left( {\theta_4
\over \theta_2}\right)^2\left( {\theta_3(z)\over \theta_3}\right)^2
\nonumber\\
Z_{11}(z) &=& -\left( {\theta_1(z)\over \theta_3}\right)^2 - \left( {\theta_1
\over \theta_3}\right)^2\left( {\theta_3(z)\over \theta_3}\right)^2
\end{eqnarray}
The contribution of the spatial twisted sectors is given by the sum of the
character of the short representation which perform the twist and an
infinite number of long representations, while in the sector twisted only
in the time direction we obtain, as one can expect, the short representation
corresponding to the identity plus long characters.\par
For example
\begin{eqnarray}
&&Z_{10} = ch^{NS}_0(l=1/2,z) - (h_3 + {\theta_4^4\over 4\eta^4})\left(
{\theta_3(z)\over \eta}\right)^2 =\nonumber\\
&&ch^{NS}_0(l=1/2,z) + \sum_i A_i q^i {q^{t-1/8}\over \eta}\left( {\theta_3(z)
\over \eta}\right)^2 \nonumber\\
&&= ch^{NS}_0(l=1/2,z) + \sum_i A_i ch^{NS}(h=t+i,l=0)
\end {eqnarray}
where $A_i,t$ are coefficients in the expansion of $h_3,\eta,\theta$ in
powers of q.
This decomposition of the partition function agrees with the previous
discussion and it explicitly shows the appearance of a number of massless
representation related to the Hirzebruck signature $\tau$.

\section{Comparison with the $SU(2)\times\r$ instanton and concluding
remarks}
\label{su2xr}
The conformal field theory based on the supersimmetrization of the
direct product of a $SU(2)$ WZNW theory at level $k$ and a Feigin-Fuchs
(F.F.)boson with $k$-dependent background charge has been widely discussed in
the literature
\cite{antoniadis,ivanov,kounnasporrati,previous,newkounnas}.
One basic feature is that
the four fermions $\lambda^0$ (the superpartner of the F.F. boson
$\partial t$) and $\lambda^i$, $i=1,2,3$ (the superpartners of the
$SU(2)_k$ currents) constitute a system of free fermions.
Another fundamental point is that if the background charge of the F.F.
boson takes the value $-i\sqrt{2\over k+2}$, the total central charge
of the theory results to be
\begin{equation}
c=c_{WZW}+c_{F.F.}+c_{free\,fermions}={3k\over k+2}+1+{6\over k+2}+2=6
\label{cc}
\end{equation}
and that the theory actually admits $N=4$ superconformal symmetry.
As emphasized in
\cite{callan} this makes {\it exact} (to all orders in $\alpha^{\prime}$)
the correspondence of this CFT with the
$\sigma$-model formulated on a certain four-dimensional target space,
a particular limit of a gravitational instanton with torsion
\cite{callan,previous}.

In \cite{previous} we focused our attention on the abstract Hodge
diamond [see (\ref{introdhodgediamond})] of the theory, and we identified
four $(1,1)$ abstract forms (giving rise to moduli of the theory)
$\Psi_{A} \sp{\um,\um}{\um,\um}^{m,{\tilde m}}(z,\bz)$.
However, as remarked by C.Kounnas in \cite{newkounnas}, more moduli
can be written. The abstract Hodge diamond shows thus a more interesting
structure; we briefly describe it. In particular it turns out that
the abstract Hodge diamond (that is intimately related, as we have
stressed throughout all the paper, to the topology of the corresponding
four-space) is sensitive to the choice of the modular invariant for the
$SU(2)_k$ subtheory.

The fundamental fields of the theory, that is the F.F. boson $t$, the
$SU(2)_k$ currents $J^i$ and the free fermions $\lambda^a$, $a=0,1,2,3$,
are normalized as follows
\footnote{We recall the expressions and normalizations of the fields
only in the left sector; those in the right one are almost identical.
As already remarked, however, non-trivial ways of merging together
left and right sector will be significative.}:
\begin{eqnarray}
\partial t(z)\partial t(w)&=&{-1\over (z-w)^2}\hskip 2pt ,\hskip 12pt
\lambda^a(z)\lambda^b(w)\,=\,-\um{\delta^{ab}\over z-w}\hskip 2pt ,\nonumber\\
J^i(z)J^j(w)&=&{k\over 2}{\delta^{ij}\over(z-w)^2}+i{\epsilon^{ijk}J^k
\over z-w}\,\, .
\label{normalizations}
\end{eqnarray}
The $SU(2)$ currents $A^i$ of the $N=4$ algebra eq.(\ref{n=4ope}) are purely
fermionic, and are therefore identical to those of the flat four-space
eq.(\ref{n=4flat}):
\begin{equation}
A^i=i(\lambda^0\lambda^i+\um\epsilon^{ijk}\lambda^j\lambda^k)\,\, .
\label{aicurrents}
\end{equation}
Denote as $\Theta_1=\left(\lambda^0+i\lambda^3\, ,\,\lambda^2
+i\lambda^1\right)$,
$\Theta_2=\left(\lambda^2-i\lambda^1\, ,\,-(\lambda^0-i\lambda^3)\right)$
the two possible doublets of fermions. The supercurrents are then
\begin{eqnarray}
&&\cG=\biggl(\partial t-\sqrt{2\over k+2}\left[J^3-2
i(\lambda^0\lambda^3-\lambda^1\lambda^2)-\partial\right]\biggr)\Theta_1+
\sqrt{2\over k+2}J^-\Theta_2\, ,\nonumber\\
&&\bar{\cal G}=\biggl(\partial t-\sqrt{2\over k+2}\left[J^3+
2i(\lambda^0\lambda^3-\lambda^1\lambda^2)-\partial\right]\biggr)\Theta^*_1+
\sqrt{2\over k+2}J^+\Theta^*_2\nonumber\, .\\
\label{supercurrentssu2xr}
\end{eqnarray}

The primary fields $:e^{i\alpha t}:$ of the F.F. theory have conformal
weight $\Delta_{\alpha}=\um\alpha(\alpha-i \sqrt{2/k+2}))$.
The primary fields corresponding to the integrable representations for
$SU(2)$ at level $k$, i.e. those of spin $l=0,\um,1,\ldots,{k\over 2}$,
have dimensions $h_l={l(l+1)\over k+2}$; we denote them as $\phi^l_m$,
$m$ being the third component (the eigenvalue of $J^3$).
It is then possible to verify that the following doublets satisfy the
basic OPEs (\ref{masslessOPE2}) so that they give rise to short
representations in the left sector:
\begin{eqnarray}
\Psi_1^{(l)}&=&e^{-\sqrt{2/k+2}(l+1)\,t}\,\phi^l_{-l}\,\Theta_1
\, ,\nonumber\\
\Psi_2^{(l)}&=&e^{-\sqrt{2/k+2}(l+1)\,t}\,\phi^l_l\,\Theta_2\, .
\label{twistedmoduli}
\end{eqnarray}
To check the OPEs (\ref{masslessOPE2}) one has to recall
the action of the third-component $J^3$ and of the raising and lowering
operators $J^{\pm}$ on the $SU(2)$ primary fields, which is:
\begin{eqnarray}
J^3(z)\phi^l_m(w)&=&m{\phi^l_m\over z-w}\, ,\nonumber\\
J^{\pm}(z)\phi^l_m(w)&=&N^l_m{\phi^l_{m\pm 1}\over z-w}\, ,
\label{su2action}
\end{eqnarray}
where $N^l_m=\sqrt{l(l+1)-m(m\pm1)}$.
It is immediate to verify the the $\Psi_A^{(l)}$ fields have
weight $\um$, as this is the weight of the fermions and
the negative weight of
$e^{-\sqrt{2/k+2}(l+1)\, t}$ cancels that of $\phi^l_m$. It is also
straightforward to write down the highest components
$\Phi_A^{(l)}$ and $\Pi_A^{(l)}$ of these short representations.

If we want to find out the numbers of abstract $(p,q)$-forms we must
specify how to put together the two sectors of the theory. The basic
constraint we have to satisfy is the modular invariance of the partition
function. In \cite{previous} we focused on the complete partition
function for the heterotic string moving on a background given by six
compactified dimensions times the four ones we are discussing. What
matters here of that discussion (see also section \ref{part}) is that we
are interested in writing the partition function {\it for fixed spin
structures} of the $(4,4)$, $c=6$ theory, $Z_{i,\bar\imath}^{(6,6)}$.
The index $i$ denotes the $SO(4)$ characters $0,v,s,\bar s$.
Since the four fermions $\lambda^a$
are effectively free, they take care of these characters. Then the other
two subtheories, namely $SU(2)_k$ and the F.F. boson,  must correspond by
themselves to modular invariant partition functions
\footnote{The situation is somehow analogous to that of
section \ref{part}. Leaving aside the fermionic spin structures,
we had there to obtain a modular invariant partition function for the orbifold
theory as in eq.(\ref{orbpart}). Here, having as basic building blocks
the affine characters $\chi_l$, a modular invariant partition function
eq.(\ref{invpart}) describes the b$SU(2)_k$ theory.}.
Of course the very
constrained part is the $SU(2)_k$ theory, for which the possible
modular invariants are well-known. These invariants depend on the
level $k$ and are related to the ADE classification \cite{adesu2}.
The invariant partition functions given in Table \ref{modinv},
where $\chi_l$, $l=0,\um,\ldots \o k2$ denotes the affine character
corresponding to the representation of spin $l$.

\begin{table}[htb]
\caption{\sl Modular invariant partition functions for $SU(2)_k$}
\label{modinv}
\begin{center}
\begin{tabular}{||c|c|c||}
\hline
Level & ADE & Partition function\\
      & (rank)    &     \\ \hline\hline
$k\geq 1$ & $A_{k+1}$ &\footnotesize{$\displaystyle{\sum_{l=0,\um,..}^{\o k2}}
\bigl|\chi_l\bigr|^2$}\\
      & $r=k+1$   &      \\
\hline
$k=4n-2$ & $D_{\o k2 +1}$ &\footnotesize{ $\displaystyle
{\sum_{l\, , int = 0}^{\o k2 -1}}
\bigl|\chi_l+\chi_{\o k2-l}\bigr|^2 + 2\bigl|\chi_{\o k4}\bigr|^2$}\\
$n\geq 1$ & $r=\o k2+1$     &      \\
\hline
$k=4n-2$ & $D_{\o k2 +2}$ &\footnotesize{ $\displaystyle
{\sum_{l\, , int. = 0}^{\o k2}}
\bigl|\chi_l\bigr|^2 + \bigl|\chi_{\o k2 +1}\bigr|^2+ \displaystyle
{\sum_{l\, ,
half\,int. =\um}^{\o k4-1}}\bigl(\chi_l^*\chi_{\o k2-l}+\chi_{\o k2-l}^*
\chi_l\bigr)$} \\
$n\geq 2$ & $r=\o k2+2$     &      \\
\hline
      &          &      \\
$k=10$ & $E_6$   &\footnotesize{ $\bigl|\chi_0 + \chi_3
\bigr|^2 + \bigl|\chi_{\o 32} +
\chi_{\o 72} \bigr|^2 + \bigl|\chi_2 + \chi_5 \bigr|^2$} \\
      & $r=6$    &      \\
\hline
      &          &\footnotesize{ $\bigl|\chi_0 + \chi_8
\bigr|^2+\bigl|\chi_2 + \chi_6 \bigr|^2
+\bigl|\chi_3 + \chi_5 \bigr|^2+\bigl|\chi_4\bigr|^2+$}      \\
$k=16$ & $E_7$  &        \\
      &          &\footnotesize{ $+\bigl[(\chi_1+\chi_7)^*\chi_4+\chi_4^*
(\chi_1+\chi_7)\bigr]$} \\
\hline
$k=28$ & $E_8$ &\footnotesize{ $\bigl|\chi_0+\chi_5+
\chi_9+\chi_{14}\bigr|^2 +
\bigl|\chi_3+\chi_6+\chi_8+\chi_{11}\bigr|^2$} \\
\hline
\end{tabular}\end{center}
\end{table}
In building the abstract $(1,1)$-forms of the theory we must take into
account the fact that indeed the F.F. boson $t$ is a function of $z$ and $\bar
z$; it is interpreted as a non-compact coordinate and we are therefore
not allowed to have independent left and right momenta: the possible
operators are just $\exp \bigl(\alpha t(z,\bar z)\bigr)$. Moreover,
we can put together a field $\phi^{l_1}_{m_1}(z)$ with
${\tilde\phi}^{l_2}_{m_2}(\bar z)$ only if, in the chosen modular
invariant partition function for the $SU(2)_k$ subtheory
\begin{equation}
\sum_{l,l^{\prime}}n_{l,l^{\prime}} \chi_l^* \chi_{l^{\prime}},
\label{invpart}
\end{equation}
the coefficient $n_{l_1, l_2}$ is different from zero.

These two considerations restrict the possible abstract $(1,1)$-forms to
be:
\begin{equation}
\Psi_{(AB)}^{(l)}\sp{\um,\um}{\um,\um}^{a,\tilde b}=
e^{(l+1)\sqrt{2/k+2}\,t(z,\bar z)} \phi^l_{m_A}(z)\Theta_A^a(z)
{\tilde\phi}^l_{m_B}(z){\tilde\Theta}_B^{\tilde b}(\bar z)
\label{unounosu2xr}
\end{equation}
for each spin $l$ such that the diagonal coefficient
$n_{l,l}$ is non-zero in the chosen $SU(2)_k$ modular invariant.
The indices $A,B$, labelling the distinct possible doublets of fermions
take the values $1,2$, so that for each $l$ there are four different
abstract $(1,1)$ forms. The third components $m_A$ are given by
$m_A=(-1)^A l_A$, see eq.(\ref{twistedmoduli}). Tildes are used to
denote fields of the right sector.

One can also see that no abstract $(1,0)$ form
$\Psi\sp{\um,0}{\um,0}^a(z,\bar z)$ can be constructed, as
the possible candidates
\begin{equation}
\Psi_{(AB)}^{(l)}\sp{\um,0}{\um,0}^a=
e^{(l+1)\sqrt{2/k+2}\,t(z,\bar z)} \phi^l_{m_A}(z)\Theta_A^a(z)
{\tilde\phi}^l_{m_B}(\bar z)
\label{unozerosu2xr}
\end{equation}
do not show the correct behaviour in the right sector.

Recall that the number of diagonal terms in each of the
possible modular invariants has a key meaning in their ADE
classification: it equals the rank $r$ of the simply-laced Lie algebra $\cG$
which corresponds to it as in Table \ref{modinv}.
We can therefore summarize the above remarks by writing the abstract
Hodge diamond of the $SU(2)_k\times\r$ theory, {\it having chosen one of
the possible partition functions for $SU(2)_k$}:
\begin{equation}
\label{hodgediamondsu2xr}
\matrix{~&~&1&~&~&\cr
{}~&0&~&0&~&\cr
1&~&4r&~&1&\cr
{}~&0&~&0&~&\cr
{}~&~&1&~&~&\cr}
\end{equation}
For example, if we are considering the theory $SU(2)_{10}\times \r$,
i.e. we set $k=10$, then we can choose three different modular
invariants, corresponding in the ADE classification to $A_{11}$, $D_7$
or $E_6$. Since the rank of these Lie algebras is respectively $11$, $7$
and $6$, the number $h^{1,1}$ of abstract $(1,1)$ forms
is very different in the
three cases; respectively we have $h^{1,1}=44$, $h^{1,1}=28$ or
$h^{1,1}=24$.

We must recall however that the four-space corresponding to the
$SU(2)\times\r$ theory is a gravitational instanton {\it with
torsion}. It admits three complex structures that are covariantly
constant with respect to connections with torsion; the same remarks
apply to its three \hika forms (such spaces were named in
\cite{previous}
generalized \hika spaces). In attributing a geometrical meaning to the
abstract Hodge diamond (\ref{hodgediamondsu2xr}) we must be aware that
the cohomology involved is that of the fiber bundle whose connections
are the torsionful connections relevant in the generalized \hika case.

The analysis in this last section makes more precise some aspects of the
treatment of the $SU(2)\times \r$ instanton that was given in
\cite{previous}; in particular we considerer there only 4 of the
possible abstract $(1,1)$ forms, i.e the short representations
$\Psi^{(0)}_{AB}\sp{\um,\um}{\um,\um}$ containing the identity  in
the $SU(2)_k$ subtheory. These are present at any level $k$, for any
chosen modular invariant. Moreover we proposed to count two abstract
$(1,0)$ forms, but as seen above it seems correct {\it not} to consider
them; the reason is the non-compactness of the coordinate $t$ which
forbids to have different momenta in left and right sector.
\vskip 0.2cm\noindent
\underline{\sl Concluding remarks}
\vskip 0.2cm
In conclusion, our analysis shows that one can study the family of
conformal field theories associated with a parametrized family of ALE
manifolds $\cM_{\zeta}$ by deforming the exactly solvable orbifold
conformal field theories $\csg$. The main open problem in this respect
has been pointed out at the end of section \ref{cft}. It relates with
a proper interpretation of the chiral ring $\c[x,y,z]/\partial W$ within
the conformal field theory. The comparison of the abstract Hodge diamond
(\ref{alehd}) of ALE theories with that of $SU(2)\times\r$ theories
(\ref{hodgediamondsu2xr}) shows that in general they are very different
theories, having a $4\times \tau$ and $16\times r$ dimensional moduli
space respectively. An unsolved problem is the geometrical
characterization of the $16\times r$-parameter family of generalized
\hika manifolds that reduces to the $SU(2)\times\r$ instanton of
\cite{callan} at a special point.
\vskip 0.2cm\begin{center}
{\bf Acknowledgements}\end{center}
\vskip 0.1cm\noindent
We are very grateful to F.Gliozzi, C.Kounnas, C.Reina for clarifying
discussions.

\end{document}